\def\etal{{\it et al.\thinspace}}
\def\mearth{{\rm\,M_\oplus}}

\documentstyle[emulateapj,psfig,apjfonts]{article}

\begin{document}

\title{High-resolution simulations of the final assembly of Earth-like
planets 1: terrestrial accretion and dynamics}

\author{Sean N. Raymond\altaffilmark{1,2}, Thomas Quinn\altaffilmark{1}, \&
Jonathan I. Lunine\altaffilmark{3}}

\altaffiltext{1}{Department of Astronomy, University of Washington, Box 351580,
Seattle, WA 98195 (raymond@astro.washington.edu)}
\altaffiltext{2}{Current address: Laboratory for Atmospheric and Space
Physics, University of Colorado, Boulder, CO 80309}
\altaffiltext{3}{Lunar and Planetary Laboratory, The University of Arizona,
Tucson, AZ 85287.}

\begin{abstract}
The final stage in the formation of terrestrial planets consists of
the accumulation of $\sim$ 1000-km ``planetary embryos'' and a swarm
of billions of 1-10 km ``planetesimals.''  During this process,
water-rich material is accreted by the terrestrial planets via impacts
of water-rich bodies from beyond roughly 2.5 AU.  We present results
from five high-resolution dynamical simulations.  These start from
1000-2000 embryos and planetesimals, roughly 5-10 times more particles
than in previous simulations.

Each simulation formed 2-4 terrestrial planets with masses between 0.4
and 2.6 Earth masses.  The eccentricities of most planets were $\sim$
0.05, lower than in previous simulations, but still higher than for
Venus, Earth and Mars.  Each planet accreted at least the Earth's
current water budget.  

We demonstrate several new aspects of the accretion process: 1) The
feeding zones of terrestrial planets change in time, widening and
moving outward.  Even in the presence of Jupiter, water-rich material
from beyond 2.5 AU is not accreted for several millions of years.  2)
Even in the absence of secular resonances, the asteroid belt is
cleared of $>$99\% of its original mass by self-scattering of bodies
into resonances with Jupiter.  3) If planetary embryos form relatively
slowly, following the models of Kokubo \& Ida, then the formation of
embryos in the asteroid belt may have been stunted by the presence of
Jupiter.  4) Self-interacting planetesimals feel dynamical friction
from other small bodies, which has important effects on the
eccentricity evolution and outcome of a simulation.

\end{abstract}


\section{Introduction}

The final stages of the formation of terrestrial planets consist of
the agglomeration of a swarm of trillions of km-sized planetesimals
into a few massive planets (see Lissauer, 1993, for a review).  Two
distinct stages in this process are usually envisioned: the formation
of planetary embryos, and their subsequent accretion into full-sized
terrestrial planets.

Runaway growth leads to the formation of Moon-to Mars-sized planetary
embryos in the inner Solar System, in a process known as ``oligarchic
growth''.  The total number of embryos is uncertain, and may range
from $\sim$ 30-50 if these are relatively massive to perhaps 500-1000
if embryos average only a lunar mass.  The timescale for embryo
formation is thought to increase with orbital radius, and decrease
with the local surface density (Kokubo \& Ida 2000, 2002).  Thus,
embryos form quickly at small orbital radii, and slower farther from
the central star.  Kokubo \& Ida (2000) found that the timescale for
embryo formation at 2.5 AU is roughly 10 million years.  However,
Goldreich \etal (2004) calculated a much shorter timescale for embryo
formation in the inner disk, $\sim 10^5$ years.

A jump in the local density may drastically reduce the timescale for
embryo formation.  Such a jump is expected immediately beyond the
``snow line'', where the temperature drops below the condensation
point of water (e.g., Stevenson \& Lunine 1988).  Beyond this jump,
the isolation masses of embryos increase, yet their formation
timescales decrease.  These massive icy embryos are thought to be the
building blocks of giant planet cores, following the
``core-accretion'' scenario for giant planet formation (Pollack \etal
1996).  Thus, embryo formation is different interior to and exterior
to the snow line -- the longest embryo formation timescales may lie in
the {\it outer} regions of the {\it inner} disk (Kokubo \& Ida 2002).

The stage of oligarchic growth is thought to end when roughly half of
the total disk mass is in the form of embryos, and half in the form of
planetesimals (Lissauer 1993).  The final assembly of terrestrial
planets consists of accretional growth of large bodies from this
swarm, on timescales of about 50 million years (e.g., Wetherill
1996).

Several authors have simulated the late-stage accretion of terrestrial
planets (e.g. Wetherill 1996; Agnor \etal 1999; Morbidelli \etal 2000;
Chambers 2001; Raymond \etal 2004, 2005a, 2005b).  Because of
computational limitations, most simulations started from only
$\sim$20-200 particles.  These simulations only included embryos in
the terrestrial region, and could not probe the full mass range of
embryos.  In addition, they inevitably neglected the planetesimal
component.

A significant physical process that requires a large number of bodies
to model is dynamical friction.  This is a damping force felt by a
large body (e.g., a planetary embryo) in a swarm of smaller bodies
(such as planetesimals).  This effect certainly plays an important
role in the final assembly of terrestrial planets (Goldreich \etal,
2004).  Indeed, several sets of dynamical simulations have formed
terrestrial planets with much higher orbital eccentricities and
inclinations than those in the Solar System (Agnor \etal 1999;
Chambers 2001; Raymond \etal 2004 -- hereafter RQL04).  Dynamical
friction could possibly reconcile this discrepancy.

Here we present results of five high-resolution simulations,
containing between 1000 and 2000 initial particles.  For the first
time, we can directly simulate a realistic number of embryos according
to various models of their formation (see section 2).  Our simulations
are designed to examine the accretion and water delivery processes in
more detail, and also to explore the dynamical effects of a larger
number of particles.  We focus our study on 1) the mass and 2) the
orbital evolution of terrestrial planets.  In addition to the
accretion of terrestrial planets, we want to understand the source of
their compositions, in particular their water contents and potential
habitability.  We address these issues in a companion paper (Raymond
\etal 2006; hereafter Paper 2).

Section 2 discusses our initial conditions, i.e. our starting
distributions of planetary embryos and planetesimals.  Section 3
explains our numerical methods.  Sections 4, 5, and 6 present the
detailed evolution of each simulation.  Section 7 concludes the paper.

\section{Initial conditions}

We choose three different sets of initial conditions for our high
resolution simulations, shown in Figure~\ref{fig:highresinit} and
summarized in Table 1.  In all cases we follow an $r^{-3/2}$ surface
density profile with total mass in solid bodies between 8.5 and 10
$\mearth$.  All protoplanets are given small initial eccentricities
($\leq 0.02$) and inclinations ($\leq 1^{\circ}$).  (Note that we use
the term ``protoplanets'' to encompass both planetary embryos and
planetesimals.  This differs from certain previous uses of the term.)

The initial water content of protoplanets is designed to reproduce the
water content of chondritic classes of meteorites (see Fig. 2 from
RQL04): inside 2 AU bodies are initially dry; outside of 2.5 AU the
have an initial water content of 5\% by mass; and between 2 and 2.5 AU
they contain 0.1\% water by mass.  The source of the water
distribution in the Solar System is a combination of heating from the
Sun (which determines the location of the snow line -- e.g. Sasselov
\& Lecar 2002) and from radioactive nuclides, which may have been
obtained from a nearby supernova early in the Sun's history (Hester
\etal 2004).  These processes are complex and not fully understood;
indeed, the location of the condensation point of water may not even
track the innermost water-rich material (Cyr \etal 1998; Kornet \etal
2004).  It is therefore unclear whether the Solar System's initial
water distribution is typical of protoplanetary disks in the Galaxy.

The starting iron contents of protoplanets are interpolated between
the values for the planets (neglecting Mercury) and chondritic classes
of meteorites, with values taken from Lodders \& Fegley (1998), as in
Raymond \etal (2005a, 2005b).  To span our range of initial
conditions, we extrapolate to values of 0.5 at 0.2 AU and 0.15 at 5
AU.

\scriptsize
\begin{deluxetable}{ccccc}
\tablewidth{0pt}
\tablecaption{Initial Conditions for 5 High Resolution Simulations}
\tablecolumns{5}
\renewcommand{\arraystretch}{.6}
\tablehead{
\colhead{Simulation} &  
\colhead{N(massive)\tablenotemark{1}} &
\colhead{N(non-int)\tablenotemark{2}} &
\colhead{$M_{TOT} (\mearth)$\tablenotemark{3}} &
\colhead{$a_{Jup} (AU)$}\tablenotemark{4}}
\startdata
 0 & 1885 & -- & 9.9 & 5.5 \\
 1a & 1038 & -- & 9.3 & 5.2 \\
 1b & 38 & 1000 & 9.3 & 5.2 \\
 2a & 1054 & -- & 8.6 & 5.2 \\
 2b & 54 & 1000 & 8.6 & 5.2 \\
\enddata
\tablenotetext{1}{Number of massive, self-interacting particles.}
\tablenotetext{2}{Number of non self-interacting particles.}
\tablenotetext{3}{Total solid mass in planetary embryos and planetesimals.}
\tablenotetext{4}{Orbital radius of Jupiter-mass giant planet.}
\end{deluxetable}
\normalsize

We begin simulation 0 in the late stages of oligarchic growth, when
planetary embryos were not yet fully formed.  The separation between
embryos is randomly chosen to lie between 0.3 and 0.6 mutual Hill
radii.  The total number of protoplanets in the simulation is 1885,
which is a factor of 5-10 more particles than previous simulations.
The surface density at 1 AU is 10 $g\,cm^{-2}$, and the total mass in
embryos is 9.9 $\mearth$, extending to 5 AU.  A Jupiter-mass giant
planet on a circular orbit is present at 5.5 AU.

In simulation 1, we follow the results of Kokubo \& Ida (2000, 2002),
who suggest that the timescale for the formation of planetary embryos
is a function of heliocentric distance.  Interior to the 3:1 resonance
with Jupiter at 2.5 AU we include embryos.  Exterior to the 3:1
resonance, we divide the total mass into 1000 ``planetesimals'' with
masses of 0.006 $\mearth$.  We attempt to probe the dynamical effects
of a swarm of planetesimals, although these are still many orders of
magnitude more massive than realistic planetesimals.  The total mass
in planetesimals and planetary embryos is 9.4 $\mearth$, with roughly
two thirds of the mass in planetesimals in the outer disk.  The
difference in total solid mass between simulations 0, 1 and 2 is not
significant, and is simply the result of random spacing of bodies.
Jupiter is included on a circular orbit at 5.2 AU -- again, this is
different than simulation 0, but not enough to cause a noticeable
effect.  We perform two simulations with the same initial conditions:
a) In simulation 1a all bodies interact with each other, b) in
simulation 1b the ``planetesimals'' past 2.5 AU are not
self-interacting.  They gravitationally interact with the embryos and
Jupiter, but not with each other.  This allows for significant
computational speedup.

In Simulation 2, we assume that planetary embryos formed all the way
out to 5 AU.  However, a significant component of the total mass is
still contained in a swarm of planetesimals which are littered
throughout the region (between 0.5 and 5 AU).  We include a total of
5.6 $\mearth$ in 54 embryos and 3 $\mearth$ in 1000 planetesimals,
which are distributed with radial distance $r$ as $r^{-1/2}$ (the
annular mass in our $r^{-3/2}$ surface density profile).  A
Jupiter-mass giant planet is included at 5.2 AU on a circular orbit.
As with simulation 1, we have run two cases -- a) one in which
planetesimals are treated in the same way as all massive bodies
(simulation 2a) and b) one in which planetesimals do not interact with
each other, but only with embryos and giant planets (simulation 2b).

Our three sets of initial conditions correspond to either different
timescales for the formation of Jupiter, or different timescales for
embryo formation.  Jupiter is constrained to have formed in the few
Myr lifetime of the gaseous component of the Solar Nebula (Brice\~no
et al 2001).  According to Kokubo \& Ida (2000, 2002), the timescale
for planetary embryos to form out to 2.5 AU is $\sim$ 10 Myr, and even
longer in the outer asteroid region (Goldreich \etal 2004 find a
timescale of $\sim 10^5$ years).  Simulation 0 contains no
embryo-sized bodies, and therefore represents a very fast timescale
for Jupiter's formation.  Simulations 1a and 1b assume embryos to have
formed out to 2.5 AU, and therefore represent a late formation for
Jupiter (in the Kokubo \& Ida model).  Simulations 2a and 2b contain
embryos out to 5 AU, and therefore represent either a very late
formation time for Jupiter, or a much faster formation time for
embryos.  Previous simulations have generally assumed embryos to exist
throughout the terrestrial zone, which is consistent with models
predicting fast growth of embryos (Goldreich \etal 2004) but not with
models predicting slower growth (Kokubo \& Ida 2000, 2002).

The most important distinction between these and previous initial
conditions is simply the scale: in the simulations from RQL04,
planetary embryo masses ranged from about 0.03 to 0.2 $\mearth$.  In
these simulations, we include bodies as small as $10^{-3} \, \mearth$
(and some even less massive ones in the inner edge of simulation 0).
The protoplanetary disks we are modeling are similar to previous
simulations, but the number of particles is larger by roughly a factor
of ten.  As shown below, our increased resolution shows several new
and interesting aspects of terrestrial accretion and water delivery.

\section{Numerical Method}

In our simulations, we include roughly 5-10 times more particles than
in previous simulations.  The difficulty in including so many
particles is that the number of operations scales strongly with the
number of particles $N$.  The computational time required for serial
algorithms such as {\tt Mercury} (Chambers 1999) scales as $N^2$,
whereas parallel codes such as {\tt pkdgrav} (Stadel 2001) can improve
the scaling to $N\,log(N)$.  One can overcome this issue in two ways:
1) simply allow simulations to run for a longer time with a serial
code such as {\tt Mercury}, or 2) use parallel machines to run
alternative algorithms such as {\tt pkdgrav} to speed up the
calculations. Here we have followed (1), and simply run simulations
for a longer time than before (we have also optimized {\tt pkdgrav}
for such calculations, which are in progress).

All simulations were evolved for at least 200 Myr, and were performed
on 2.7 GHz desktop PCs using {\tt Mercury} (Chambers 1999).
Simulation 0 required 16 consecutive months of integration time.
Simulation 1a and 2a took 4-5 months each.  Simulations 1b and 2b took
only 2-3 months of CPU time because most particles were not
self-interacting, thereby greatly speeding up {\tt Mercury}'s $N^2$
force algorithm.  We use a timestep of 6 days, to sample the innermost
body in our simulation twenty times per orbit.  Collisions are treated
as inelastic mergers.  Energy is conserved to better than 1 part in
10$^3$ in all cases.

\section{Simulation 0: 1885 initial particles}

Figure~\ref{fig:aet2000} shows six snapshots in time in the evolution
of simulation 0.  Embryos quickly become excited by their own mutual
gravitation as well as that of the Jupiter-mass planet at 5.5 AU (not
shown in plot).  Objects in several distinct mean motion resonances
have been excited by the 0.1 Myr snapshot and can be seen in
Fig.~\ref{fig:aet2000}: the 3:1 resonance at 2.64 AU, 2:1 resonance at
3.46 AU, and the 5:3 resonance at 3.91 AU.  All material exterior to
the 3:2 resonance at 4.2 AU is quickly removed from the system via
collisions with and ejections by the giant planet, which we refer to
as Jupiter, for simplicity.  Once eccentricities are large enough for
particles' orbits to cross, bodies begin to grow via accretionary
collisions.  This has started to happen in the inner disk by 1 Myr.
These large bodies tend to have small eccentricities and inclinations,
due to the dissipative effects of dynamical friction (equipartition of
energy causes the smaller bodies to be more dynamically heated).

Accretion proceeds fastest in the the inner disk and moves outward, as
can be seen by the presence of several large bodies inside 2 AU after
10 Myr.  In the outer disk, collisions are much less frequent because
of the slower dynamical timescales and decreasing density of material
due to Jupiter clearing the region via ejections.  However, the mean
eccentricity of bodies in the outer disk is quite high, enabling those
which are not ejected by Jupiter to have their orbits move inward in
time.  After 30 Myr, the inner disk is composed primarily of a few
large bodies, which have begun to accrete water-rich material from the
outer disk.  Larger bodies have started to form out to 2.5-3 AU, but
their eccentricities are large.  In time, several of these outer large
bodies are incorporated into the three final terrestrial planets.  We
name the surviving planets $a$, $b$, and $c$ (innermost to outermost).
The properties of all the planets formed in our five high-resolution
simulations are listed in Table 2.  Plots like Fig.~\ref{fig:aet2000}
of the evolution of each simulation are shown later in the paper.

\scriptsize
\begin{deluxetable}{ccccccccc}
\tablewidth{0pt}
\tablecaption{Properties of {\bf(potentially habitable)} planets formed\tablenotemark{1}}
\tablecolumns{9}
\renewcommand{\arraystretch}{.6}
\tablehead{
\colhead{Simulation} &  
\colhead{Planet} &  
\colhead{$a (AU)$} &
\colhead{$\bar{e}$\tablenotemark{2}} &
\colhead{$\bar{i} (^{\circ})$\tablenotemark{3}} &
\colhead{M$(\mearth)$} &
\colhead{W.M.F.} &
\colhead{N(oceans)\tablenotemark{4}} &
\colhead{Fe M.F.}}

\startdata
 0 & a & 0.55 & 0.05 & 2.8 & 1.54 & $2.6\times 10^{-3}$ & 15 & 0.32\\
   & \bf b &  \bf 0.98 &  \bf 0.04 & \bf 2.4 & \bf 2.04 & ${ \bf 8.4\times 10^{-3}}$ & \bf 66 & \bf 0.28\\
   & c & 1.93 & 0.06 & 4.6 & 0.95 & $9.1\times 10^{-3}$ & 33 & 0.28\\
1a & a & 0.58 & 0.05 & 2.7 & 0.93 & $8.3\times 10^{-3}$ & 30 & 0.31\\
   & \bf b & \bf 1.09 & \bf 0.07 & \bf 4.1 & \bf 0.78 & ${\bf 5.5\times 10^{-3}}$ & \bf 17 & \bf 0.30\\
   & c & 1.54 & 0.04 & 2.6 & 1.62 & $1.2\times 10^{-2}$ & 75 & 0.26\\
1b & a & 0.52 & 0.06 &  8.9  & 0.60 & $7.2\times 10^{-3}$ & 17 & 0.31\\
   & \bf b & \bf 1.12 & \bf 0.05 & \bf 3.5  & \bf 2.29 & ${\bf 6.7\times 10^{-3}}$ & \bf 60 & \bf 0.29\\
   & c & 1.95 & 0.09 & 9.7 & 0.41 & $3.8\times 10^{-3}$ & 6 & 0.28\\
2a & a & 0.55 & 0.08 & 2.6 & 1.31 & $9.3\times 10^{-4}$ & 5 & 0.33\\
   & \bf b & \bf 0.94 &\bf 0.13 &\bf 3.4  & \bf 0.87 & ${\bf 8.6\times 10^{-3}}$  &\bf  29 & 0.29\\
   & \bf c & \bf 1.39 & \bf 0.11 &\bf 2.4  & \bf 1.46 &  ${\bf 6\times 10^{-3}}$ & \bf 34 & 0.29\\
   & d & 2.19 & 0.08 & 8.8  &  1.08 & $1.8\times 10^{-2}$ & 75 & 0.24\\
2b & a & 0.61 & 0.18 & 13.1 &  2.60 & $7.1\times 10^{-3}$ & 71 & 0.30\\
   & b & 1.72 & 0.17 & 0.5 & 1.63 & $2.0\times 10^{-2}$ & 126 & 0.22\\

\\
Mercury\tablenotemark{5} & & 0.39 & 0.19 & 7.0 & 0.06 & $1\times 10^{-5}$ & 0 & 0.68\\
Venus &   & 0.72 & 0.03 & 3.4 & 0.82 & $5\times 10^{-4}$ & 1.5 & 0.33\\
\bf Earth & & \bf 1.0 & \bf 0.03 & \bf 0.0 & \bf 1.0 &  ${\bf 1 \times
  10^{-3}}$ & \bf 4 & \bf 0.34\\
Mars &    & 1.52 & 0.08 & 1.9 & 0.11 & $2\times 10^{-4}$ & 0.1 & 0.29\\
\enddata
\tablenotetext{1}{Planets are defined to be $> 0.2 \mearth$.  Shown in
bold are bodies in the habitable zone, defined to be between 0.9 and
 1.4 AU.  This is slightly wider than the habitable zone of Kasting
 \etal (1993).}
\tablenotetext{2}{Mean eccentricity averaged during the last 50 Myr of
 the simulation.}
\tablenotetext{3}{Mean inclination averaged during the last 50 Myr of
 the simulation.}
\tablenotetext{4}{Amount of planetary water in units of Earth oceans,
 where 1 ocean = $1.5 \times 10^{24}\, (\approx 2.6 \times 10^{-4} \mearth)$.}
\tablenotetext{5}{Orbital values for the Solar system planets are 3
  Myr averaged values from Quinn \etal (1991).  Water contents are
  from Morbidelli \etal (2000).  Iron values are from Lodders \&
  Fegley (1998).  The Earth's water content lies between 1 and 10
  oceans -- here we assume a value of 4 oceans.  See text for discussion.} 
\end{deluxetable}
\normalsize

\subsection{Mass Evolution of simulation 0 planets}

In RQL04, the number of constituent particles that ended up in a given
Earth-mass planet was typically 30-50.  The three planets in
simulation 0 formed from a much larger number of bodies: planet $a$
(0.55 AU) from 500 protoplanets via 87 accretionary collisions, planet
$b$ (0.98 AU) from 457 protoplanets via 98 collisions, and planet $c$
(1.93 AU) from 174 protoplanets via 47 collisions.

Figure~\ref{fig:frac-a2000} shows the ``feeding zone'' of each planet,
the starting location of all bodies which were incorporated into that
planet.  Planet $a$ contains 8\% (5\%) of material from past 2 AU (2.5
AU), planet $b$ contains 25\% (17\%) from past 2 AU (2.5 AU), and
planet $c$ contains 37\% (18\%) from past 2 AU (2.5 AU).  The feeding
zone of planet $a$ is much narrower than those of planets $b$ and $c$,
and is concentrated in the inner terrestrial zone.  However, the
feeding zones of the planets change in time.

Figure~\ref{fig:m-t2000} shows the mass of each of the three final
planets as a function of time.  Planets $a$ and $b$ start to grow
quickly, but planet $c$ starts later because of the longer dynamical
timescales in the outer terrestrial zone.  We expect planet $a$ to
start growing faster than planet $b$ for the same reason, but this is
not the case.  This is because the accretion seed of planet $b$
originated very close to the seed of planet $a$, and was in the same
region of rapid accretion at early times.  Each planet grows quickly
in the first few million years by accreting local material, but this
rate flattens off within 20-30 million years.  This happens fastest
for planet $a$ because the accretion in the inner disk is shortest,
and most available material has been consumed within that time.  The
later stages of growth are characterized by a step-wise pattern
representing a smaller number of larger-scale collisions with other
``oligarchs'', which have also cleared out all bodies in their feeding
zones.  Such large, late impacts are reminiscent of the Moon-forming
impact (e.g., Canup \& Asphaug, 2001), which is thought to have
occurred at low relative velocity and an oblique impact angle.

There exists a constraint from measured Hf-W ratios that both the Moon
and the Earth's core were formed by $t\, \approx$ 30 Myr (Kleine
\etal, 2002; Yin \etal, 2002), suggesting that the Earth was at least
50\% of its final mass by that time.  Each of the three planets in
simulation 0 reached half of its final mass within 10-20 Myr.

Figure~\ref{fig:mloc-t2000} shows the bodies accreted by each of the
surviving planets from simulation 0 (planet $a$ in green, planet $b$
in blue, planet $c$ in red) as a function of time.  The y axis
represents the starting semimajor axis of each body, or in the case of
objects which had accreted other bodies, the mass-weighted starting
semimajor axis of the agglomeration.  The relative sizes of accreted
objects is also depicted.  In the early stages of accretion, each
planet accretes material from the inner disk, and the feeding zones of
planets $a$ and $b$ overlap.  In time, the planets clear their feeding
zones of available bodies, and increase their mass and gravitational
focusing factors.

Bodies in the outer terrestrial region and asteroid belt have their
eccentricities pumped up via resonant excitation and secular forcing
by Jupiter, as well as mutual close encounters.  Many of these bodies
are accreted by the planets.  Until there exist large bodies with
sufficient mass to scatter smaller bodies, eccentricities in a given
region remain low and the region is dynamically isolated.  The
accretion of more distant bodies at later times reflects the dynamical
change in these regions, from isolated groups composed of many small
bodies to interacting groups dominated by a few large bodies.  In
effect, this change marks the end of oligarchic growth in a given
region.  The values from Fig.~\ref{fig:mloc-t2000} are slightly longer
than for simulations of oligarchic growth (Kokubo \& Ida 2000).  This
is because we are measuring the time for a body to grow and be
accreted by planets at large distances rather than the formation time
of planetary embryos, though the dynamical consequences are similar.
For example, planets $a$, $b$, and $c$ began accreting material from
2.5 AU and beyond after $\sim$ 10 Myr, as compared with a 10 Myr
oligarchic growth timescale from Kokubo \& Ida (2000).

Figure~\ref{fig:mloc-t2000} shows that it is not until after about
$10^7$ years that water-rich bodies are accreted by the planets.  At
this late time, the feeding zones of all three planets are the same,
and extend beyond the snow line.  The vast majority of the water
content of each planet originates between 2.5 and 4 AU.  Material
between 2-2.5 AU does not contain enough water, and the dynamical
lifetimes of bodies past 3.5-4 AU are very short.  This is discussed
in more detail in Paper 2.

Figure~\ref{fig:imass-t2000} shows the mass of bodies accreted by each
of the surviving planets from simulation 0 as a function of time.  The
size of each body is proportional to its mass$^{1/3}$, and therefore
represents its relative physical size.  Bodies of a range of sizes are
accreted by each planet throughout the simulation, although large
impacts tend to happen later in the simulation.  Some of these large
impacts may form large satellites (moons).  This is discussed in
detail in Paper 2.

The radial time dependence of accretion seen in
Fig.~\ref{fig:mloc-t2000} is reflected in the positions of ejected
bodies.  Figure~\ref{fig:ejec-loc2000} shows the starting semimajor
axes as a function of time of all bodies that were either ejected
(blue) or accreted (red) by Jupiter.  These bodies have had their
eccentricities increased and had close encounters with Jupiter, which
has either devoured them or scattered them from the system.  During
the first part of the simulation, only bodies past 4 AU are destroyed
by Jupiter.  But after $\sim 10^7$ years, the initial location of
ejected bodies is a scatter plot, signaling the end of the isolation
of accretion zones.  Bodies from throughout the terrestrial region are
scattered into interstellar space via close approaches to Jupiter.

In a disk of purely massless or non self-interacting particles, there
would be no change in the trend like the one in
Fig.~\ref{fig:ejec-loc2000} at $10^7$ years.  Rather, through secular
forcing, bodies from the outer asteroid belt would have close
encounters with Jupiter and be ejected.  The zone cleared out by
Jupiter would slowly widen in time, asymptotically reaching the point
where the forced eccentricity could not induce a close enough approach
to Jupiter.  Additional zones would be cleared out via mean motion
resonances.  However, there would be no abrupt changes in the zones
from which particles are ejected or accreted.  This illustrates an
important effect of the self-gravity of the protoplanetary disk.

\subsection{Eccentricity Evolution}

Previous simulations of terrestrial accretion have not succeeded in
reproducing the very low eccentricities of Earth, Mars, and Venus
(Agnor \etal 1999; Chambers 2001; RQL04).  The final eccentricities of
the three planets at the end of the integration are 0.03, 0.09, and
0.07 for planets $a$, $b$, and $c$, respectively.  However, these are
instantaneous values not necessarily representative of the long-term
evolution of the system.  The mean eccentricities for planets $a$,
$b$, and $c$ from 100-200 Myr are 0.048, 0.039, and 0.057.

Figure~\ref{fig:eall2000} shows the eccentricity evolution of each
planet from simulation 0 through time.  During the first 20 Myr, while
these planets were accreting material at a very high rate, their
eccentricities remain relatively low, below 0.05-0.10.  Collisions
tend to circularize orbits, because they are most likely to happen
when one body is at aphelion and the other at perihelion.  In the
period from 20-80 Myr, the planets' eccentricities increase and
fluctuate wildly due to numerous close encounters with other massive
bodies, which may increase eccentricities, but a slower rate of
collisions.  After this time, the planets' eccentricities are damped
through interactions with the remaining small bodies in the system.
The planets' eccentricities reach very low values at $t \sim 150$ Myr.
The mean eccentricities of the three planets from 150-170 Myr are
0.033 for planet $a$, 0.026 for planet $b$, and 0.057 for planet $c$.

At about 173 Myr, there is a jump in the eccentricities of the two
inner planets, from about 0.03 to 0.05.  This occurred due to the
scattering of a Moon-sized (0.02 $\mearth$) body from the asteroid
belt into the inner terrestrial zone.  Figure~\ref{fig:a-t2000} shows
the semimajor axes vs time of all surviving bodies for the last 50 Myr
of the integration.  Jupiter and the terrestrial planets are labeled
by their final masses.  Several bodies are ejected between 150-170
Myr, shown as vertical spikes as their semimajor axes increase to
infinity.  Beginning at $t \approx 160$ Myr, two bodies in the
asteroid belt begin to interact strongly and have several close
encounters.  At $t = 173$ Myr, the smaller, roughly lunar-mass body is
scattered by the larger, Mars-sized body into the terrestrial region.
Its eccentricity is large enough that over the next twenty million
years, this body has several close encounters with all three of the
terrestrial planets, causing a large jump in their eccentricities.  In
time, the body is scattered outward and collides with Jupiter at $t =
193$ Myr.  Incidentally, the roughly Mars-mass scattering body is the
largest object in the asteroid belt at this time, but is ejected from
the system at $t = 198$ Myr.

It is interesting to note in Fig.~\ref{fig:eall2000} how closely the
eccentricity of the inner two planets track each other in time.
Throughout the simulation, most sharp jumps are seen in the
eccentricity evolution of both bodies, and sometimes in the outer
planet as well.  By examining the planets' longitudes of perihelion,
we have determined that planets $a$ and $b$ are in a secular
resonance.  Figure~\ref{fig:plam0} shows a histogram of the relative
orientation of the two planets' orbits.  The relative orientation is
measured by the quantity $\Lambda$, the difference in the two planets'
longitudes of perihelion, normalized to lie between 0 and 180 degrees
(Barnes \& Quinn 2004).  The sharp spike seen at $\Lambda \,
\approx \, 120 ^{\circ}$ with the tail to higher $\Lambda$ values is the
signature of secular resonance.  It indicates that the orbits of
planets $a$ and $b$ are librating about anti-alignment, with an
amplitude of about 60$^\circ$.  The two planets' eccentricities are
out of phase -- planet $a$'s eccentricity is low when planet $b$'s is
high, and vice versa.  An interesting case from simulation 2a
involving habitability will be discussed in section 8.

We explore two measures of the total dynamical excitement of the
system as a function of time.  The mass weighted eccentricity of all
bodies is simply

\begin{equation}
MWE = \frac{\sum_j m_j e_j}{\sum_j m_j},
\end{equation}

\noindent where we sum over all surviving bodies $j$ (excluding
Jupiter).  The angular momentum deficit measures the deviation of a
set of orbits from perfect, coplanar circular orbits.  This makes
intuitive sense because the orbital angular momentum correlates
roughly with the perihelion distance, so a circular orbit has the
maximum angular momentum for a given semimajor axis.  The angular
momentum deficit includes both the inclination and eccentricity of
bodies in its formulation, and is defined by Laskar (1997) as

\begin{equation}
S_d = \frac{\sum_{j} m_j \sqrt{a_j} \left(1 - cos (i_j)
\sqrt{1-e_j^2}\right)} {\sum_j m_j \sqrt{a_j}},
\end{equation}

\noindent where, in this case,  $i_j$ refers to a body's inclination
with respect to the plane of Jupiter.

Figure~\ref{fig:mwe-angmom2000} plots the mass-weighted eccentricity
and angular momentum deficit of all bodies through the course of the
simulation.  The two measures of dynamical excitement track each other
fairly well through the simulation, although the angular momentum
deficit contains many more rapid fluctuations in time.  This is
because $S_d$ is more sensitive than the $MWE$ to small bodies, which
may attain very high eccentricities and inclinations through close
encounters.  Each spike in $S_d$ is due to a single, high-eccentricity
(or high-inclination) particle which does not persist in this excited
state.  Rather, such a particle quickly falls into the Sun or is
ejected from the system.  For example, the four ejection events
between 150-170 Myr seen in Fig.~\ref{fig:a-t2000} can be correlated
with small spikes in the angular momentum deficit in
Fig.~\ref{fig:mwe-angmom2000}.  Short term variations in the
mass-weighted eccentricity are due to a combination of close
encounters which pump up eccentricities, and destruction of high-$e$
bodies via ejections or collisions, which decreases the mean
eccentricity.  In addition, both the mass weighted eccentricity and
angular momentum deficit exhibit the jump at 173 Myr associated with
the close encounter shown in Fig~\ref{fig:a-t2000}.

The mass weighted eccentricity of the four terrestrial planets in the
Solar System and the largest asteroid, Ceres, is 0.037.  The mean for
the final 20 Myr of simulation 0 is 0.069, but for the low
eccentricity period between 150 and 170 Myr, it is 0.046.  The
terrestrial planets in this simulation have significantly lower
eccentricities than in most previous simulations (e.g. Chambers 2001;
RQL04).  However, they are still larger than those in the Solar
System.

Using 3 million year average values from Quinn \etal (1991), the
angular momentum deficit of the Solar System terrestrial planets and
Ceres is 0.0017, much smaller than $S_d$ in
Fig.~\ref{fig:mwe-angmom2000}.  However, the value from simulation 0
is dominated by the few small bodies that remain in the asteroid belt
on relatively high eccentricity and inclination orbits.  If we
restrict ourselves to only the three massive terrestrial planets, the
value of $S_d$ decreases sharply to roughly 0.0037 for the last 20 Myr
of the simulation.  In the period of very low eccentricities, from
about 150-170 Myr, $S_d$ averaged 0.0026, only 50\% higher than for
the recent, time-averaged Solar System.  If we include only the
terrestrial planets in the calculation, then $S_d$ remains unchanged
at 0.0017.  However, if we include only Venus, Earth, and Mars, $S_d$
drops to 0.0013.  This indicates that the planets in simulation 0 are
slightly more dynamically excited than the Solar System, but not by a
large amount.  

\subsection{Evolution of the asteroid belt}

The evolution of orbits in the asteroid belt, defined to be exterior
to 2.2 AU (roughly the 4:1 resonance with Jupiter), differs from that
in the terrestrial region, as no large bodies remain after 200 Myr.
The top panel of Figure~\ref{fig:astbelt2000} shows the most massive
body in the asteroid belt from simulation 0 through time.  Bodies up
to 0.26 $\mearth$ may exist in the asteroid belt.  However, these
bodies are not able to survive in this region, and are removed by
scattering from either Jupiter or other asteroid belt bodies.  Many
embryo-sized bodies did not accrete in the asteroid belt, but rather
formed closer to the Sun and were scattered outward, where their
eccentricities were damped causing them to remain on orbits in the
asteroid belt.  Accretion did take place in the asteroid region in
simulation 0, but only to a limited extent.  Nine bodies originating
in the region reached masses of 0.05 $\mearth$, and four reached 0.1
$\mearth$.  The most massive locally-formed body has 0.2 $\mearth$.
Accretion was primarily confined to the inner belt, as no bodies
larger than 0.1 $\mearth$ formed past 3 AU.

Many previous authors (e.g. Agnor \etal, 1999; Morbidelli \etal, 2000;
Chambers 2001) assumed that all of the mass in the asteroid belt was
in the form of $\geq$Mars-sized planetary embryos.  If these embryos
form via oligarchic growth, separated by 5-10 mutual Hill radii, these
would have masses in the asteroid belt between about 0.08 and 0.17
$\mearth$.  By starting our simulation from smaller bodies, we see
that only a handful of such objects could actually form in the region
on the relevant timescales (although we have not included the effects
of very small bodies, which may be significant -- Goldreich \etal
2004).  Our results suggest that initial conditions containing embryos
throughout the asteroid belt may not be realistic.  This has important
implications for the water delivery process, and is discussed in Paper
2.

The middle panel of Fig.~\ref{fig:astbelt2000} shows the evolution of
the total mass in the asteroid belt.  Our initial conditions included
objects out to 5 AU.  Virtually all particles exterior to the 3:2
Jupiter resonance at 4.2 AU were removed in the first few thousand
years, causing the immediate steep drop from 4.9 $\mearth$ to 4.2
$\mearth$ in the first 10 thousand years.  After the initial drop,
mass is continuously lost from the asteroid belt.  Most objects are
either scattered into the terrestrial region to be incorporated into
the terrestrial planets, or undergo a close encounter with Jupiter and
are ejected from the system.  The total mass ejected over the course
of the simulation is 3.08 $\mearth$ out of 9.9 $\mearth$ of total
solid mass.  In addition, 0.66 $\mearth$ collided with Jupiter, and
0.02 $\mearth$ hit the Sun.

The asteroid belt is almost completely cleared of material, even
though Jupiter is on a circular orbit.  After 200 Myr, only five
bodies remain in the asteroid belt, four of which are clustered around
3.1 AU (see Fig.~\ref{fig:aet2000}), corresponding to the outer main
belt in our Solar System, which is dominated by C-type asteroids.  The
total mass of these five surviving bodies in simulation 0's asteroid
belt is 0.07 $\mearth$, roughly 1.5\% of the total mass.  We continued
the simulation to 1 Gyr, at which time only one 0.03 $\mearth$ body
remained in the asteroid belt, comprising only 0.6\% of the initial
mass.  Models of the primordial surface density of the asteroid belt
suggest that the asteroid belt in our Solar System has been depleted
by a factor of $10^2 - 10^3$ (e.g., Weidenschilling 1977).  Our
results imply that such clearing is a natural byproduct of the
evolution of the disk.  As suggested by Morbidelli \etal (2000), the
asteroid belt may be cleared simply by self-scattering of bodies onto
unstable orbits such as resonances with Jupiter.  We show that this
can occur even with a very calm dynamical environment, with Jupiter on
a circular orbit.  In this situation, Jupiter's orbit does not precess
(indeed, precession is meaningless with a circular orbit), so there
are no secular resonances in the asteroid belt (such as the $\nu_6$
resonance with Saturn at 2.1 AU).  Tsiganis \etal (2005) suggest that
the giant planets did not acquire appreciable eccentricities until
about 600 Myr after the start of planet formation.  Thus, even in this
framework, the asteroid belt may have been rapidly cleared by the
self-gravity of the disk.

The bottom panel of Fig.~\ref{fig:astbelt2000} shows the mass-weighted
eccentricity through time of all bodies in the asteroid belt.  A
comparison with Fig.~\ref{fig:mwe-angmom2000} shows that bodies in the
asteroid belt have significantly higher eccentricities than those in
the inner terrestrial region throughout the simulation.  This is
simply because bodies in the asteroid belt feel stronger perturbations
than bodies closer to the Sun, due to Jupiter's proximity.  In other
words, the magnitude of the forced eccentricity for any particle in
the asteroid belt is higher than for one in the terrestrial region.
In time, the amplitude of short term oscillations in the $MWE$
increases as the number of bodies decreases.

\section{Simulations 1a and 1b: 1038 initial particles}

Simulations 1a and 1b both start with 38 planetary embryos to 2.5 AU,
and 1000 0.006 $\mearth$ ``planetesimals'' between 2.5 and 5 AU (see
Fig.~\ref{fig:highresinit}).  The starting positions of all embryos
and planetesimals are identical in simulations 1a and 1b.  The only
difference between the two simulations is how planetesimals are
treated.  In simulation 1a, planetesimals are massive bodies which
interact gravitationally with all other bodies in the simulation.  In
simulation 1b, however, planetesimals feel the gravitational presence
of the embryos and Jupiter, but not each other's presence.  In
addition, these non-interacting planetesimals may not collide with
each other.  Any differences in the outcome of the two simulations
therefore has to do with the properties of these particles.

The reasons for running a simulation which includes non
self-interacting particles are several.  Physically, we chose to run
simulation 1b to test the importance of the self-gravity among small
bodies in the protoplanetary disk (included in simulation 1a, but not
in 1b).  The effects are surprisingly important, as described below.
In addition, the computational expense of including test particles is
much less than for self-interacting particles, and scales with the
number of particles $N$ as $N$ instead of $N^2$.  Therefore, many more
such particles may be included in a simulation.  However, the
limitations of such simulations must be understood.

Figures~\ref{fig:aetfe1a} and~\ref{fig:aetfe1b} show six snapshots in
the evolution of simulations 1a and 1b, respectively.  One difference
between the two simulations that is immediately evident is the shape
of mean-motion resonances with Jupiter in the asteroid belt.  In
simulation 1a, the vertical structure of these resonances is washed
out within 1 Myr, as resonant planetesimals excited by Jupiter excite
other planetesimals in turn, and effectively spread out the resonance.
In simulation 1b, however, the resonant structure is more confined and
lasts longer, because of the lack of planetesimal-planetesimal
encounters.  The shape of the 3:2 resonance with Jupiter in simulation
1b is still intact after 1 Myr, and can be seen in
Fig.~\ref{fig:aetfe1b}.  In both simulations, a swarm of surviving
planetesimals appears to sweep into the terrestrial region at between
10 and 30 Myr, damping the eccentricities of large bodies.  These
planetesimals also deliver water to the surviving planets.  In time,
the numbers of planetesimals dwindle, but at different rates for the
two simulations.  This has important effects on the final
eccentricities of the surviving planets, and is discussed below.

There are several similarities in the final planetary systems from
simulations 1a and 1b.  The positions of the two inner planets in each
simulation are comparable (see Table 2 for details), although the
masses of these planets differ significantly.  The total mass in the
three surviving planets from each simulation is nearly identical, 3.3
$\mearth$.  The total water content of these planets is also similar,
although the planets in simulation 1a are slightly more water-rich,
containing a total of 132 oceans of water as compared with 83 oceans
contained in the planets of simulation 1b.

There exist significant differences between simulations 1a and 1b.
Figure~\ref{fig:massvs1} shows the growth of the surviving terrestrial
planets in simulations 1a (top panel) and 1b (bottom panel).  It is
clear that planets form faster in simulation 1a -- they reach a
significant fraction of their final mass at shorter times.  In
addition, the final large accretion event for each planet in
simulation 1a occurs before $t \approx$ 60 Myr, while in simulation 1b
this happens between roughly 70-90 Myr.

The main difference between the two simulations is the timing of the
destruction of particles.  Gravitational self-scattering of
planetesimals in simulation 1a decreases the residence time of these
bodies in the asteroid region compared with simulation 1b, and
scatters them onto unstable orbits.  Eventually, these bodies either
have a close encounter with Jupiter and are ejected, or collide with a
larger protoplanet.  In addition, planetesimal collisions may occur,
further decreasing the number of particles, although this effect is
relatively minor.

The overall effects of this self-gravity are similar to those of the
mass of asteroid region planetesimals explored in RQL04.  In RQL04, we
showed that more massive planetesimals cause the terrestrial planets
to be more massive, more water-rich, to form more quickly, and to be
fewer in number.  Here we see that the self-gravity of planetesimals
accelerates planet formation and moderately increases the water
content of the terrestrial planets.  The final planets from
simulations 1a and 1b both contain roughly the same amount of mass.
We can apply this to our result from RQL04, that a larger planetesimal
mass causes the formation of a smaller number of more massive planets.
That effect must be due to the amount of mass in the asteroid region
in the ``case (i)'' simulations from RQL04, and not just dynamical
self-stirring.

The top panel of Figure~\ref{fig:numvs1} shows the evolution of the
number of surviving particles in simulations 1a (blue) and 1b (red).
The number of planetesimals drops much more rapidly in simulation 1a,
and all are destroyed by $t \approx 140$ Myr.  The total mass in each
simulation that either collides with Jupiter or is ejected is very
similar, roughly 6 $\mearth$ of the 9.3 starting $\mearth$.  Most of
this is in the form of planetesimals, although some embryos are
ejected.  The evolution of the embryos is similar for the two cases.
This is not surprising as embryos reflect the evolution of the inner
disk, which starts with no planetesimals.

The bottom panel of Fig.~\ref{fig:numvs1} shows the mass-weighted
eccentricity of all surviving particles from simulations 1a and 1b as
a function of time.  Eccentricities in both simulations grow rapidly
at the start.  Simulation 1b reaches higher eccentricities in the
heavy accretion phase before 50 Myr.  This is because planetesimals
cannot damp their own eccentricities, which are elevated from a
distance by secular and resonant perturbations from the embryos and
Jupiter.  In simulation 1a, collisions and dynamical stirring among
planetesimals damp these eccentricities.  In time, the eccentricities
of both systems decrease, and flatten off at similar levels, with
mass-weighted eccentricities of 0.05-0.07.  Eccentricities are barely
damped at all in either simulation after 100 Myr, as the number of
total bodies has been reduced to just a few.

We may have run into a resolution limit, as no planetesimals remain at
the end of simulation 1a or 1b.  The only refuge of such bodies is the
asteroid region, separated from both the terrestrial planets and
Jupiter.  The dynamical effects of small bodies on the planets are
important in terms of damping eccentricities and inclinations.  It is
not clear from our simulations whether the vast majority of small
bodies are necessarily destroyed within 200 Myr, meaning that even
higher-resolution simulations are useless.  A more complex treatment
of terrestrial accretion could generate additional particles during
each collision.  This remains as an avenue for future study.

It is clear from the 10 Myr and 30 Myr panels of
Figs.~\ref{fig:aetfe1a} and~\ref{fig:aetfe1b} that planetary embryos
do exist in the asteroid region in both simulations 1a and 1b, even
though the planetesimals in simulation 1b cannot accrete.  These
embryos did not form {\it in situ}, but were fully formed at the start
of the simulation, interior to 2.5 AU, and were scattered into the
asteroid belt in a process similar to ``orbital repulsion'' (Kokubo \&
Ida 1995).  The embryos' eccentricities were increased by
gravitational stirring, and they were scattered beyond 2.5 AU.
Dynamical friction acting on the large bodies damped their
eccentricities without greatly altering their semimajor axes, thereby
moving them out into the asteroid region.  In this area, the orbits of
embryos can be stable on 100 Myr timescales.  Indeed, in simulation
1a, one embryo formed at 1.8 AU, quickly grew to 0.35 $\mearth$, and
was scattered into the asteroid region.  It survived there for 200 Myr
until being ejected at 211 Myr.

Few embryos formed in the asteroid belt in simulation 1a.  Although
accretion in the asteroid region did occur, only a handful of bodies
accreted multiple other planetesimals, and only five accumulations
contained four or more planetesimals.  Indeed, the largest
accumulation contained 8 planetesimals totaling only 0.05 $\mearth$,
comparable in size to the smallest embryos in the inner part of the
disk.  Embryos in the 2-2.5 AU region, separated by 5-10 mutual Hill
radii, are typically 0.1-0.2 $\mearth$.  So, even the largest
planetesimal accumulation was a factor of 2-3 smaller than its
isolation mass.  No large embryos formed in the asteroid belt in
simulation 1a, although a few embryos did form in the asteroid region
in simulation 0, as discussed above.  We conclude that accretion can
happen in the asteroid belt, but only to a limited degree, forming a
smaller number of less massive embryos than assumed in most previous
accretion simulations (e.g. Morbidelli \etal, 2000).

\section{Simulations 2a and 2b: 1054 initial particles}

\begin{figure}
\centerline{\psfig{file=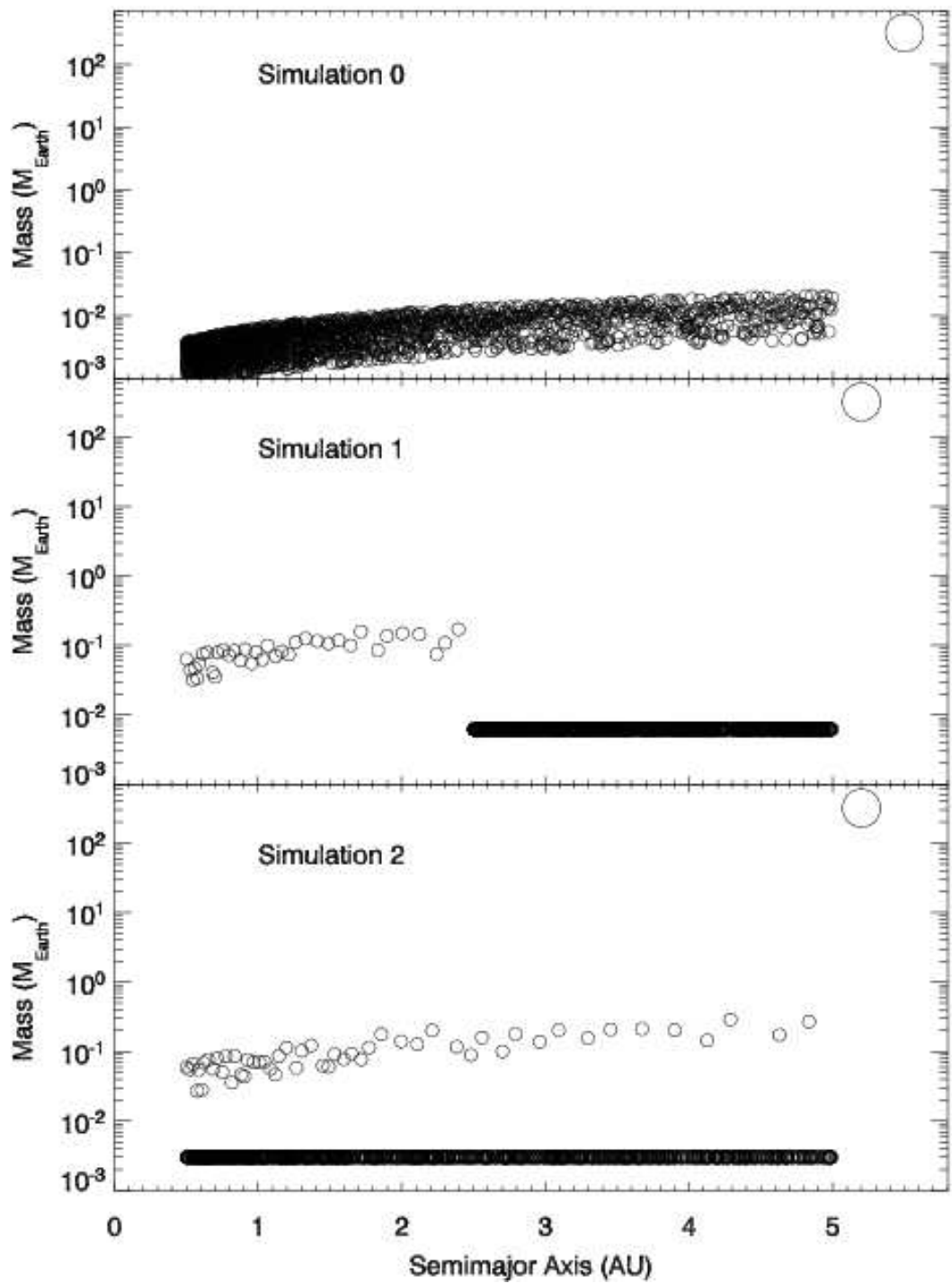,width=15cm}}
\caption{Our three sets of initial conditions for high-resolution
  simulations.  The large circles at 5.5 and 5.2 AU are Jupiter-mass
  giant planets.  The horizontal lines in simulations 1 and 2
  represent 1000 planetesimals, which are distributed with orbital
  radius $r$ as $r^{-1/2}$, corresponding to the annular mass in a
  disk with an $r^{-3/2}$ surface density profile.}
\label{fig:highresinit}
\end{figure}

\begin{figure}
\centerline{\psfig{file=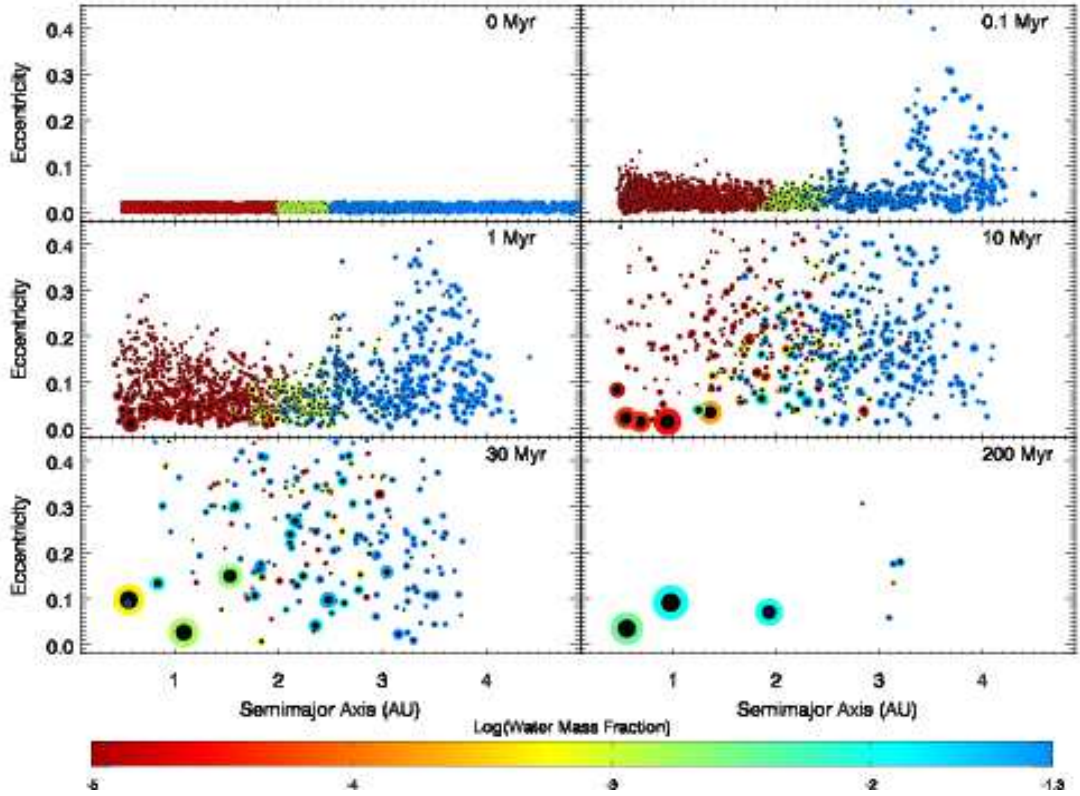,width=15cm}}
\caption{Six snapshots in time from simulation 0, with 1885 initial
  particles.  The size of each body corresponds to its relative
  physical size (i.e. its mass $M^{1/3}$), but is not to scale on the
  x axis.  The color of each particle represents its water content,
  and the dark inner circle represents the relative size of its iron
  core.  There is a Jupiter-mass planet at 5.5 AU on a circular orbit
  (not shown).}
\label{fig:aet2000}
\end{figure}

\begin{figure}
\centerline{\psfig{file=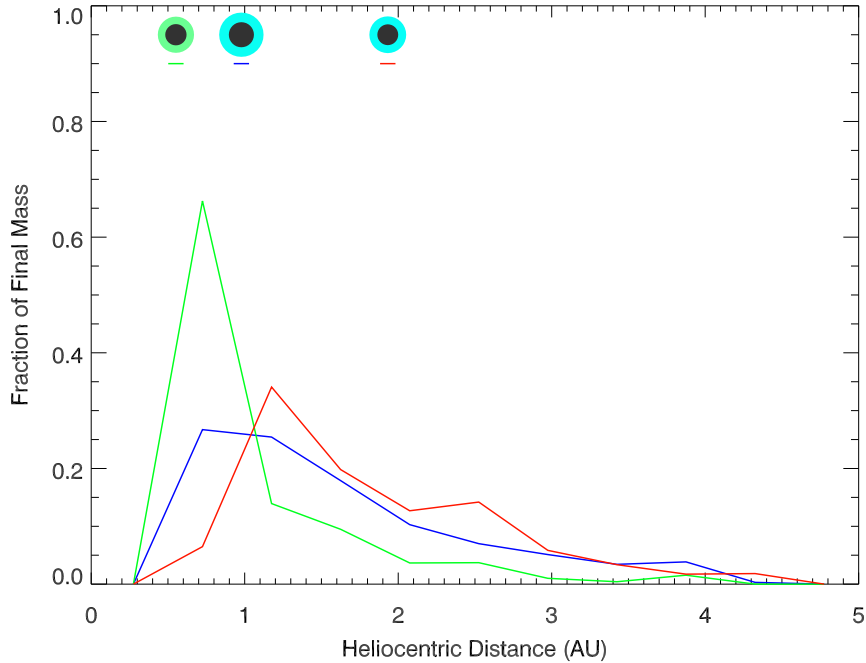,width=13cm}}
\caption{The feeding zones of the three surviving massive planets from
simulation 0 (see Table 2).  The y axis represent the fraction of each
planet's final mass that started the simulation in a given 0.45 AU
wide bin.  The final configuration of the planets is shown at top of
the figure (as in the 200 Myr panel from Fig.~\ref{fig:aet2000}).  The
color of each curve refers to the horizontal colored line below each
planet.}
\label{fig:frac-a2000}
\end{figure}

\begin{figure}
\centerline{\psfig{file=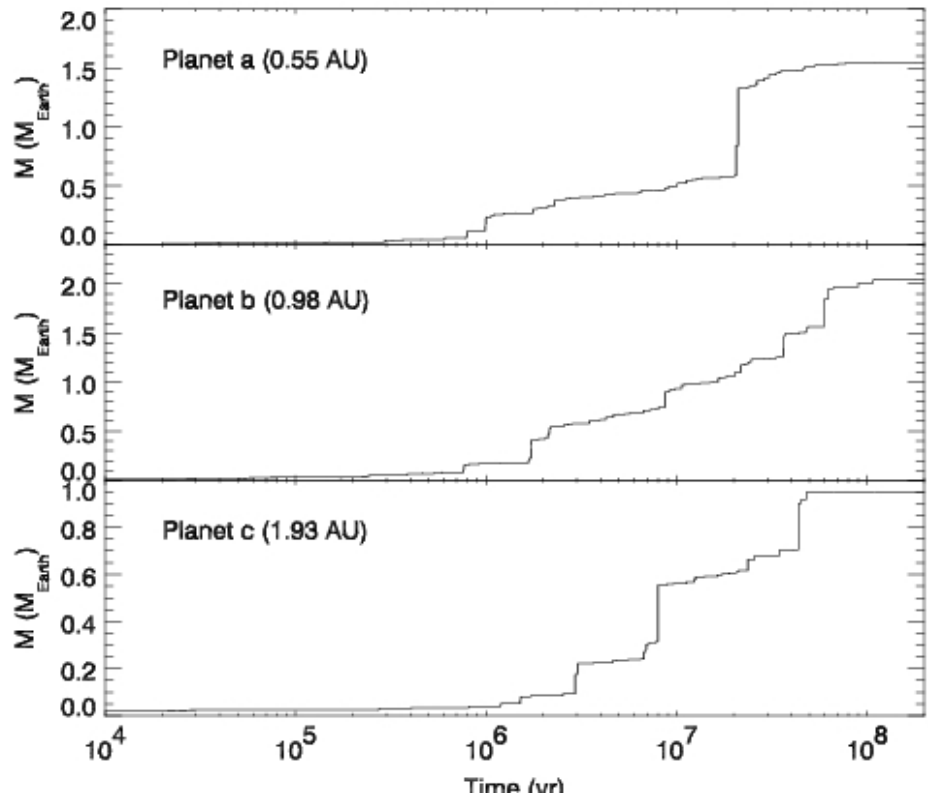,width=13cm}}
\caption{Mass vs time for the three surviving planets from simulation
0.}
\label{fig:m-t2000}
\end{figure}

In simulations 1a and 1b we assumed that embryos formed only out to
2.5 AU, but in simulations 2a and 2b we assume that they formed all
the way out to 5 AU.  This corresponds to either a fast growth of
embryos, as advocated by Goldreich \etal (2004), or perhaps a very
slow growth of Jupiter.  Simulations 2a and 2b have identical starting
conditions (see Fig.~\ref{fig:highresinit}), with 54 planetary embryos
throughout the terrestrial region, embedded in a disk of 1000
``planetesimals'' of 0.003 $\mearth$ each.  Roughly two thirds of the
mass is in the form of embryos, and one third in planetesimals.  The
only difference between simulations 2a and 2b is the way in which the
planetesimals are treated: in simulation 2b the planetesimals do not
feel each other's presence, but in simulation 2a all particles are
fully self-interacting.

Interestingly, the initial conditions for simulations 1a and 1b are
reasonable if embryos form via the Goldreich \etal (2004) model, but
are unrealistic if they follow the results of Kokubo \& Ida (2000,
2002).  The oligarchic growth phase is thought to end when the total
mass in planetesimals and embryos is comparable (e.g., Lissauer
1993). Our initial conditions therefore represent a time after the
completion of oligarchic growth.  However, Kokubo \& Ida's model of
oligarchic growth suggests that there may not be enough time for
embryos to form in the outer terrestrial zone, in the asteroid region
beyond 2-2.5 AU.  Indeed, analysis of simulations 0 and 1a suggests
that, if Jupiter was fully formed before embryos formed in the
asteroid region, then perhaps five to ten embryos could have formed in
this region.  These would be sub-isolation mass objects, with masses
of $\sim 0.05 \mearth$ rather than the $0.1-0.2 \mearth$ embryos in
Fig.~\ref{fig:highresinit}.  In Kokubo \& Ida (2000)'s model, embryos
at 2.5 AU take roughly 10 Myr to form.  The lifetime of gaseous disks
around other stars is $\leq$ 10 Myr (Brice\~no \etal 2001),
constraining the formation time of gas giant planets.  Embryos form
more slowly at larger orbital radii, so this model predicts that
oligarchic growth in the asteroid region must have occurred in the
presence of the giant planet, stunting the formation of embryos.

Figures~\ref{fig:aetfe2a} and~\ref{fig:aetfe2b} show snapshots in time
of simulations 2a and 2b, respectively.  The evolution of the two
simulations is qualitatively similar in the first few Myr, but the
outcomes are drastically different.  Simulation 2a forms four
terrestrial planets, including two in the habitable zone.  Simulation
2b, in contrast, forms only two planets, one interior to and one
exterior to the habitable zone.

Interactions among planetesimals appear to be very important to the
dynamics of the system.  The outcome of simulation 2b is very
surprising in that the final terrestrial planets have very high
eccentricities ($e \sim 0.2$), not expected for a $>$1000 particle
simulation.  Eccentricities in simulation 2a are slightly higher than
in simulations 0, 1a and 1b, but are only about half of those in
simulation 2b ($e \sim 0.1$).  The final angular momentum deficit of
the planets in simulation 2a is one third of that for simulation 2b.

The early evolution of the simulations is not surprising -- bodies are
excited by resonances and mutual encounters.  The shapes of resonances
are quickly washed out by the interactions among embryos.  Dynamical
friction between the embryos and planetesimals damps the
eccentricities of the embryos and increases that of the planetesimals.
This process appears to be stronger in simulation 2a than in
simulation 2b.  In each of the final three panels of
Figs.~\ref{fig:aetfe2a} and~\ref{fig:aetfe2b}, large bodies in
simulation 2a have significantly smaller eccentricities than those in
simulation 2b.  The feeding zone of each planet is affected by its
eccentricity; a larger eccentricity implies a larger radial variation
during a body's orbit, causing it to intercept more material than for
a circular orbit (Levison \& Agnor 2003).  Indeed, the final planets
in simulation 2b have high eccentricities and large masses.

Figures~\ref{fig:evol2a} and~\ref{fig:evol2b} show the time evolution
of several aspects of simulations 2a and 2b, respectively.  The top
panels show the growth of the two final planets.  The middle panels
show the number of surviving bodies, both embryos and planetesimals,
through time.  The bottom panels show the mass-weighted eccentricity
of all surviving bodies as a function of time.  Note that the scale of
the y axis for the bottom panels is different in the two figures.

The formation timescales are similar in simulations 2a and 2b.  In
both cases most accretion takes place in the first 30 Myr.  The
planets reach a significant fraction of their final masses by the end
of rapid accretion.  After this time, the planets accrete (often
water-rich) planetesimals as well as possibly undergoing a large
impact with another embryo.

In both simulations, the evolution of the mass-weighted eccentricity
begins with the same rapid increase and bumpy profile seen in
simulation 0 (Fig.~\ref{fig:mwe-angmom2000}).  The eccentricities in
simulation 1a in this early stage are slightly lower than those in
simulation 2b.  Eccentricities decrease in both simulations after $t
\approx 40$ Myr, after which their evolutions diverge.  In both
simulations, the decrease is followed by a sharp increase.  However,
the magnitude and duration of the increase are drastically different.
In simulation 1a, the mass-weighted eccentricity reaches 0.2 at $t$ =
50 Myr, then begins a slow decline for the rest of the simulation.  In
simulation 2b, the eccentricity reaches values of almost 0.4 at $t
\approx 60$ Myr, and does not flatten out until 100 Myr.  When it does
flatten out, it does so at a much higher level than in the other
simulations, about 0.2.  The middle panel of Fig~\ref{fig:evol2b}
shows that the increase in eccentricity at 50 Myr corresponds to the
time when only a very few large bodies remain in the system, at the
end of the rapid accretion phase.  Indeed, only four embryo-sized
bodies survive to $t = 45$ Myr.  As seen in the upper panel of
Fig~\ref{fig:evol2b}, two of these bodies are accreted by the
surviving planets at 76 Myr and 127 Myr.  However, the number of
embryos at this time in simulation 2a is comparable.  The differences
lie in the evolution of the planetesimals.

During the time of very high eccentricities in simulation 2b, between
roughly 50 and 100 Myr, only a few large bodies remain.  Less than 100
planetesimals remain at the beginning of this period, and their
destruction rate is accelerated at $t \sim 50$ Myr, such that all are
ejected by $t \approx 75$ Myr.  The increase in eccentricities is due
to the lack of dissipative forces, i.e. no small bodies remain to
provide dynamical friction.  The same acceleration in the destruction
rate of planetesimals is seen in simulation 1a at $t \sim 55$ Myr.
However, the curve flattens out, and several tens of planetesimals
remain after the sharp increase in eccentricities at 50 Myr.  Thus,
large bodies in simulation 1a continue to feel dynamical friction,
while those in simulation 1b do not.

Why do self-interacting planetesimals provide more dissipation than
non self-interacting ones?  In simulation 2a, eccentricities reached
by the planetesimals are much smaller than in simulation 2b.  This
implies that interactions among planetesimals damp their
eccentricities, i.e. planetesimals can feel dynamical friction from
other planetesimals.  By damping planetesimals' excitations, more are
on low-inclination orbits likely to be encountered by embryos,
increasing the damping.  In addition, by staying on lower-excitation
orbits, the lifetime of self-interacting planetesimals is longer.
Unfortunately, the computational time needed to integrate the orbits
of $N$ self-interacting bodies scales as $N^2$, while non
self-interacting bodies only scale as $N$.

It is interesting that self-interacting planetesimals survive longer
in simulation 2a than simulation 2b, but non self-interacting
planetesimals survive longer in simulation 1b than simulation 1a.  In
simulations 1a, planetesimals are primarily destroyed by ejection
after being scattered by other planetesimals onto unstable resonances
with Jupiter in the asteroid region.  Non self-interacting particles
in unstable resonances are cleared out quickly, but nearby orbits are
not perturbed into instability.  In contrast, the main destruction
mechanism of planetesimals in simulations 2a and 2b is scattering by
embryos onto unstable orbits.  The eccentricities of non
self-interacting planetesimals increase more rapidly than
self-interacting planetesimals, so they are destroyed more quickly.

When all the planetesimals are destroyed, simulations proceed in the
same fashion as previous, low resolution simulations which formed
planets with relatively large eccentricities (e.g. Chambers 2001;
RQL04).  Secular perturbations and close encounters among the few
remaining bodies excite their eccentricities.  The feeding zones of
these eccentric embryos are enlarged, and bodies are accreted until
the system stabilizes with a small number of relatively massive,
eccentric terrestrial planets.  This occurs in simulation 2b, as
planetesimals are destroyed before the end of accretion.  In
simulation 2a, however, some planetesimals survive through the end of
accretion, keeping eccentricities low and planetary feeding zones
narrow.  The surviving planets in simulations 2a and 2b reflect this
important difference in the evolution of their planetesimal
populations.

A problem with our simulations is that the number of particles can
only decrease.  Realistic collisions between large bodies create a
swarm of debris, the details of which depend on the collision velocity
and angle (Agnor \& Asphaug, 2004).  Interactions between large bodies
and this collisional debris tends to circularize the orbits of large
bodies via dynamical friction (Goldreich \etal 2004).  The reason for
the high eccentricities in simulation 2b is the rapid destruction of
all the small bodies.  We included enough planetesimals to ``resolve''
the early parts of the simulation, but the planetesimals were all
destroyed quickly.  This problem may be solved in two ways, by either
starting the simulation with even more particles than included here,
by using self-interacting planetesimals (as in simulation 2a), or by
creating debris particles during collisions.  It would be very
interesting to self-consistently create debris particles during
oblique or high-velocity collisions, but this is left to future work.

\section{Conclusions}

We have run five simulations of terrestrial accretion with
unprecedented resolution.  The planets that formed were qualitatively
similar to those formed in previous simulations, such as RQL04.  The
mean eccentricities of our planets are smaller than in previous
simulations, averaging about 0.05 vs. 0.1, but they remain larger than
those of Venus, Earth, and Mars.  This may be due to the resolution of
our simulations, as the number of small bodies dwindles at late times.
Perhaps a higher resolution simulation will further damp the
terrestrial planet eccentricities.  Future simulations may also
include additional dissipative effects, such as accounting for
dynamical friction from collisional debris.

\begin{figure}
\centerline{\psfig{file=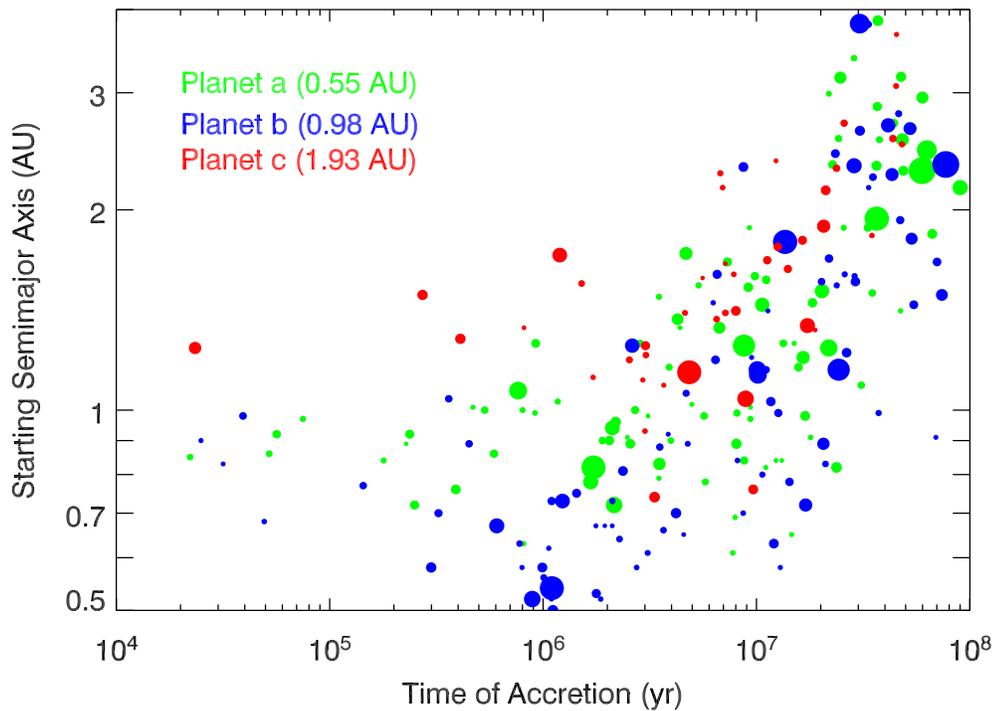,width=15cm}}
\caption{The timing of accretion of bodies from different initial
  locations for simulation 0.  All bodies depicted in green were
  accreted by planet $a$, all blue bodies were accreted by planet $b$,
  and all red bodies were accreted by planet $c$.  The relative size
  of each circle indicates its actual relative size.  Impactors which
  had accreted other bodies are given their mass-weighted starting
  positions.}
\label{fig:mloc-t2000}
\end{figure}

\begin{figure}
\centerline{\psfig{file=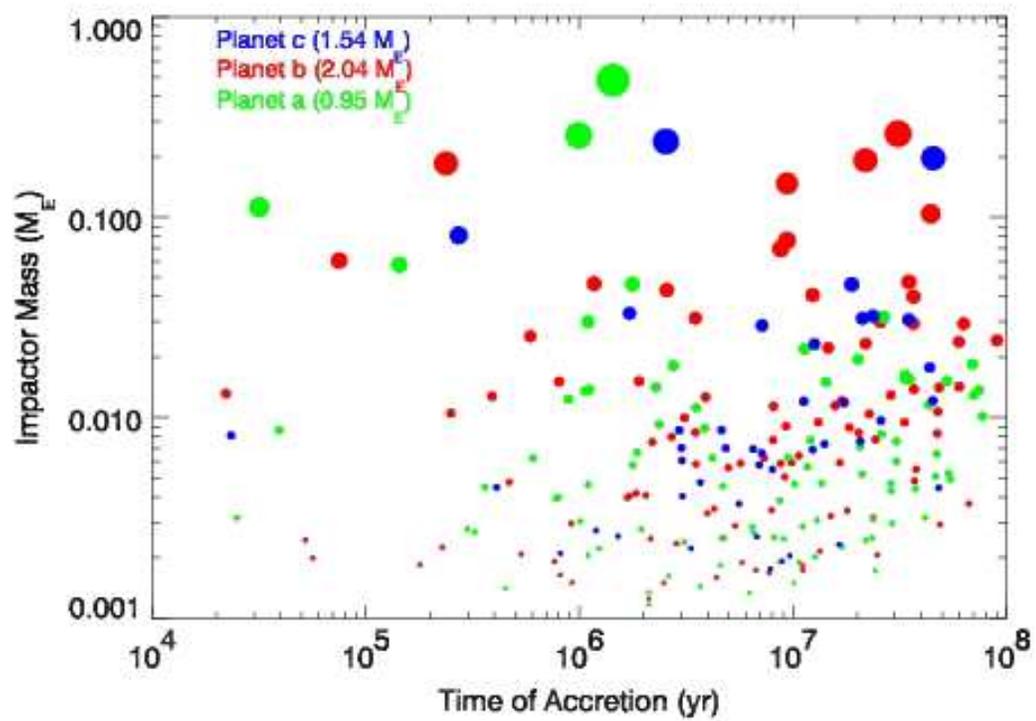,width=15cm}}
\caption{Impactor mass vs time for all bodies accreted by the
  surviving planets in simulation 0.  As in Fig.~\ref{fig:mloc-t2000},
  all bodies depicted in green were accreted by planet $a$, etc. The
  size of each body is proportional to its mass$^{1/3}$, and
  represents its relative physical size.}
\label{fig:imass-t2000}
\end{figure}

\begin{figure}
\centerline{\psfig{file=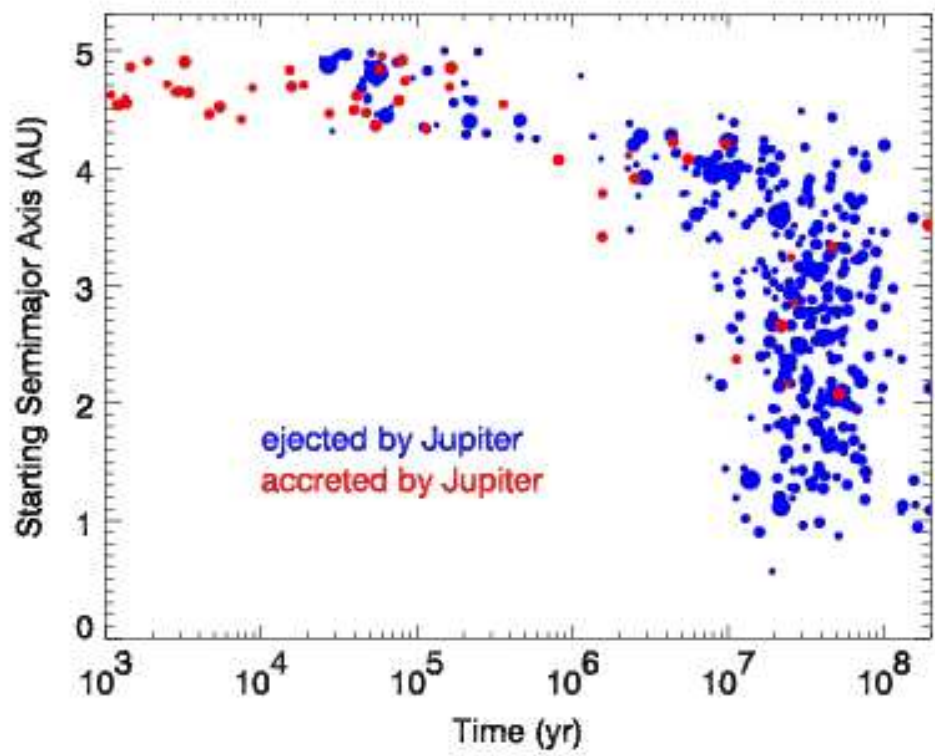,width=15cm}}
\caption{The timing of the ejection and accretion by Jupiter of bodies
  from different starting locations.  Note the change in the
  ``ejection zone'' at $t \approx 10 Myr$.}
\label{fig:ejec-loc2000}
\end{figure}

\begin{figure}
\centerline{\psfig{file=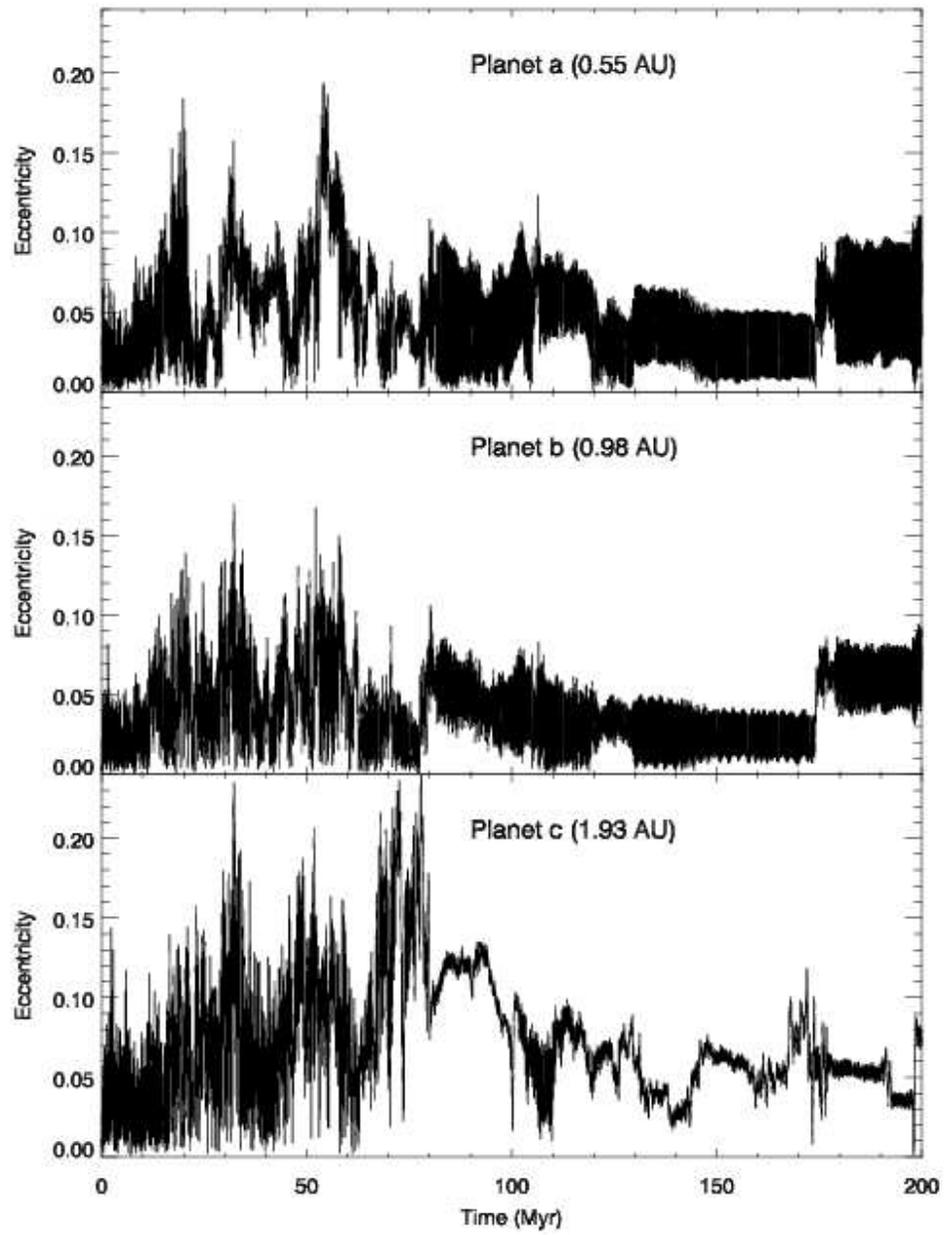,width=13cm}}
\caption{Eccentricity vs time for the three surviving massive
  planets.  See Table 2 for details.}
\label{fig:eall2000}
\end{figure}

\begin{figure} 
\psfig{file=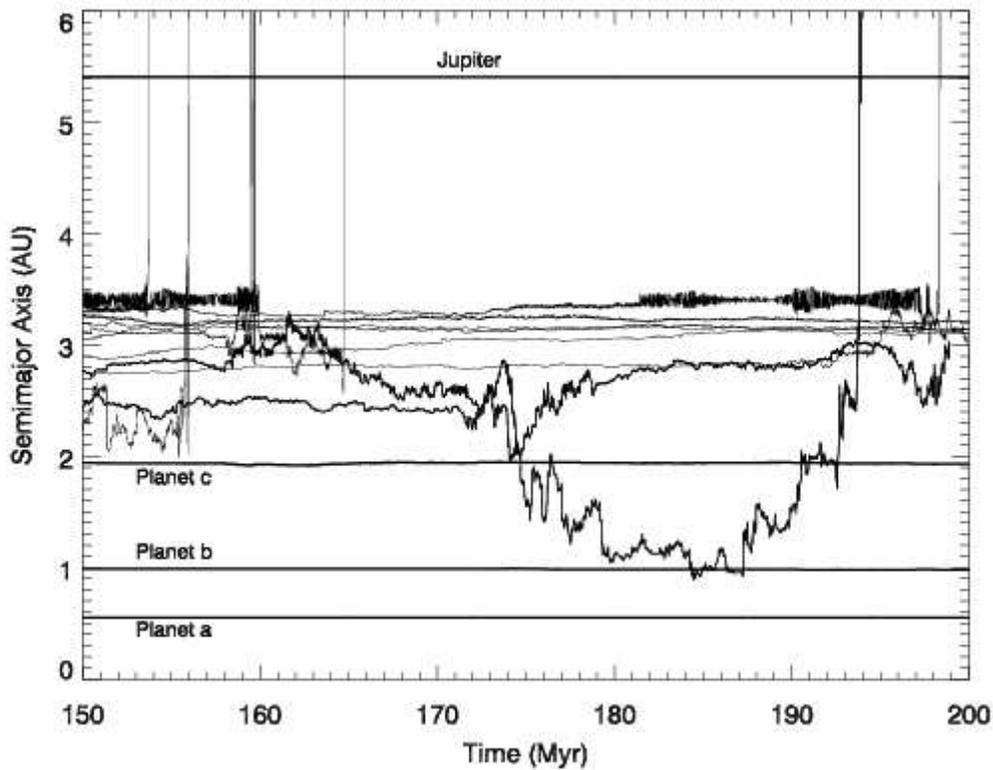,width=15cm} 
\caption{Semimajor axes of all bodies from simulation 0 for the time
period from 150-200 Myr.  The three surviving terrestrial planets and
  Jupiter are in bold and are labeled.  Vertical spikes indicate a
  particle's ejection from the system.  Notice the encounter between a
  stray body and the terrestrial planets from about 173 to 193 Myr,
  which had a large effect on the eccentricities of the surviving
  planets (see Fig.~\ref{fig:eall2000}).  The body's eccentricity was
  large enough ($> 0.4$) that its orbit crossed that of planet $a$.}
\label{fig:a-t2000}
\end{figure}

\begin{figure}
\centerline{\psfig{file=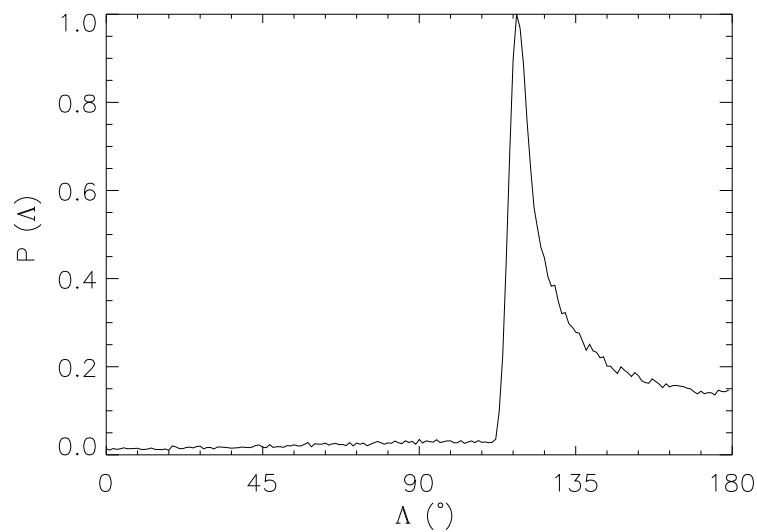,width=11cm}}
\caption{Normalized histogram of the relative orientation of the orbits of
planets $a$ and $b$ from simulation 0 over one billion years.
$\Lambda$ represents the difference in the two planets' longitudes of
perihelion, normalized to lie between 0 and 180 degrees (Barnes \&
Quinn 2004).  The spike at $\Lambda \approx \, 120 ^{\circ}$ and tail
to higher $\Lambda$ values is reminiscent of a harmonic oscillator.
It indicates that the two planets are locked in secular resonance,
librating about anti-alignment with an amplitude of about 60 degrees.}
\label{fig:plam0}
\end{figure}

\begin{figure}
\centerline{\psfig{file=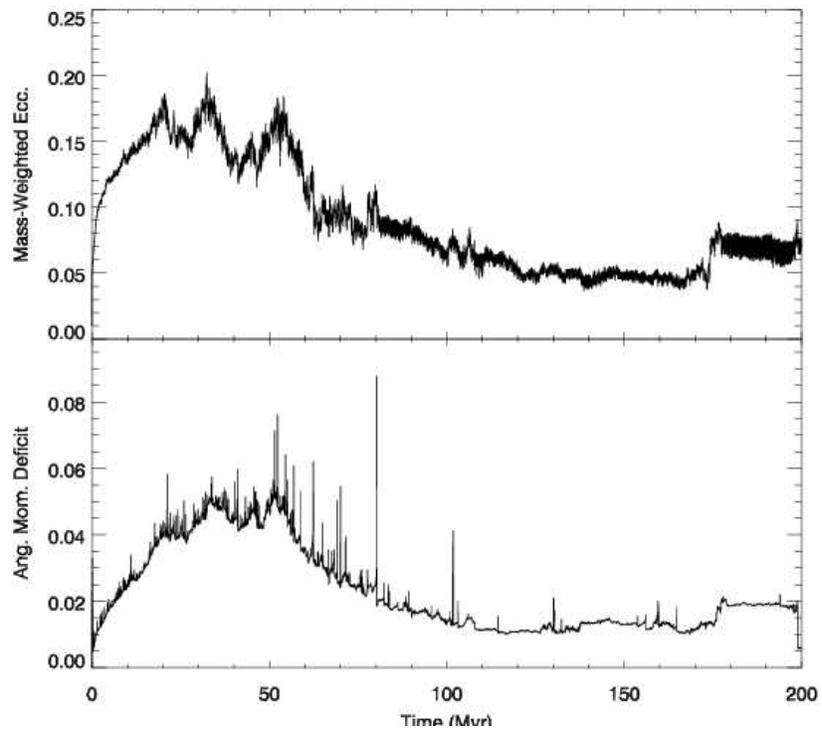,width=11cm}}
\caption{ Top panel -- Mass weighted eccentricity vs time for all
bodies from simulation 0, excluding Jupiter.  Bottom Panel -- Angular
momentum deficit vs time for all bodies from simulation 0, also
excluding Jupiter.  These quantities are defined in Eqs. 1 and 2.}
\label{fig:mwe-angmom2000}
\end{figure}

\begin{figure}
\centerline{\psfig{file=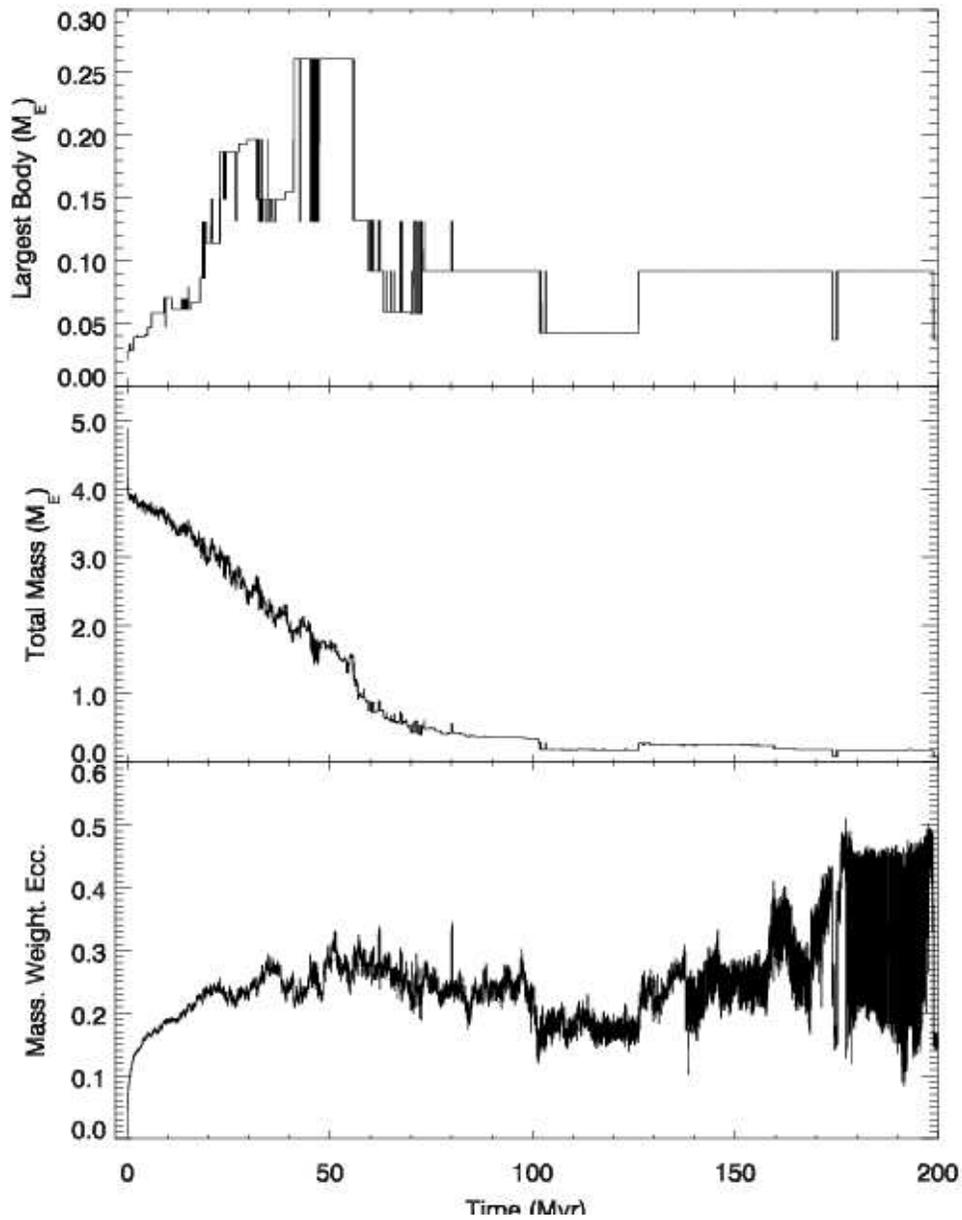,width=13cm}}
\caption{Evolution of the asteroid belt (defined as $2.2<a<5.2$ AU) in
time for simulation 0.  \underline{Top} -- the most massive body in
the asteroid belt through time.  \underline{Middle} -- total mass in
the asteroid belt as a function of time.  \underline{Bottom} --
Mass-weighted eccentricity of all bodies in the asteroid belt over
time.}
\label{fig:astbelt2000}
\end{figure}

\begin{figure}
\centerline{\psfig{file=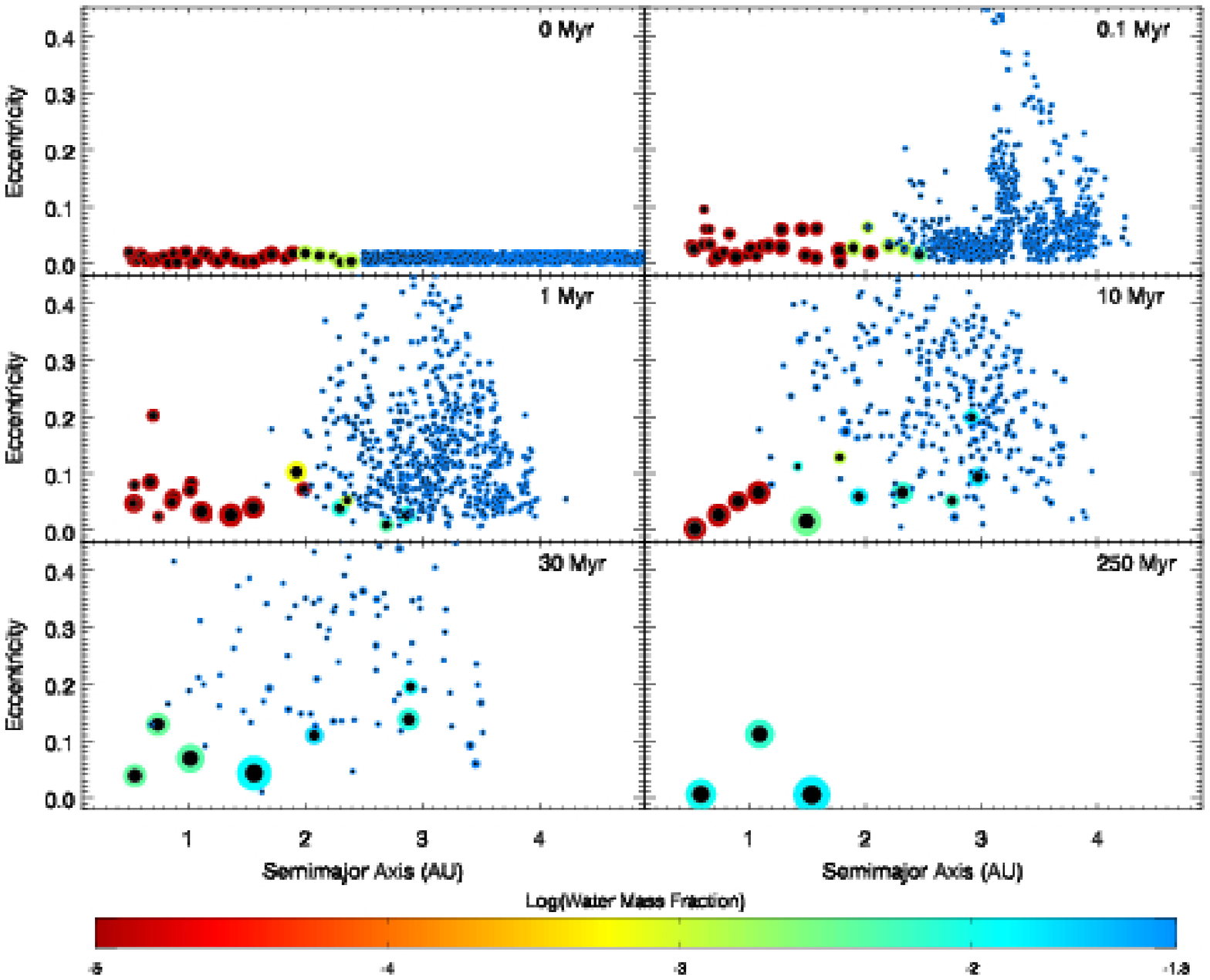,width=12cm}}
\caption{Six snapshots in the evolution of simulation 1a, with 1038
  initial particles, all self-interacting.}
\label{fig:aetfe1a}
\end{figure}

\begin{figure}
\centerline{\psfig{file=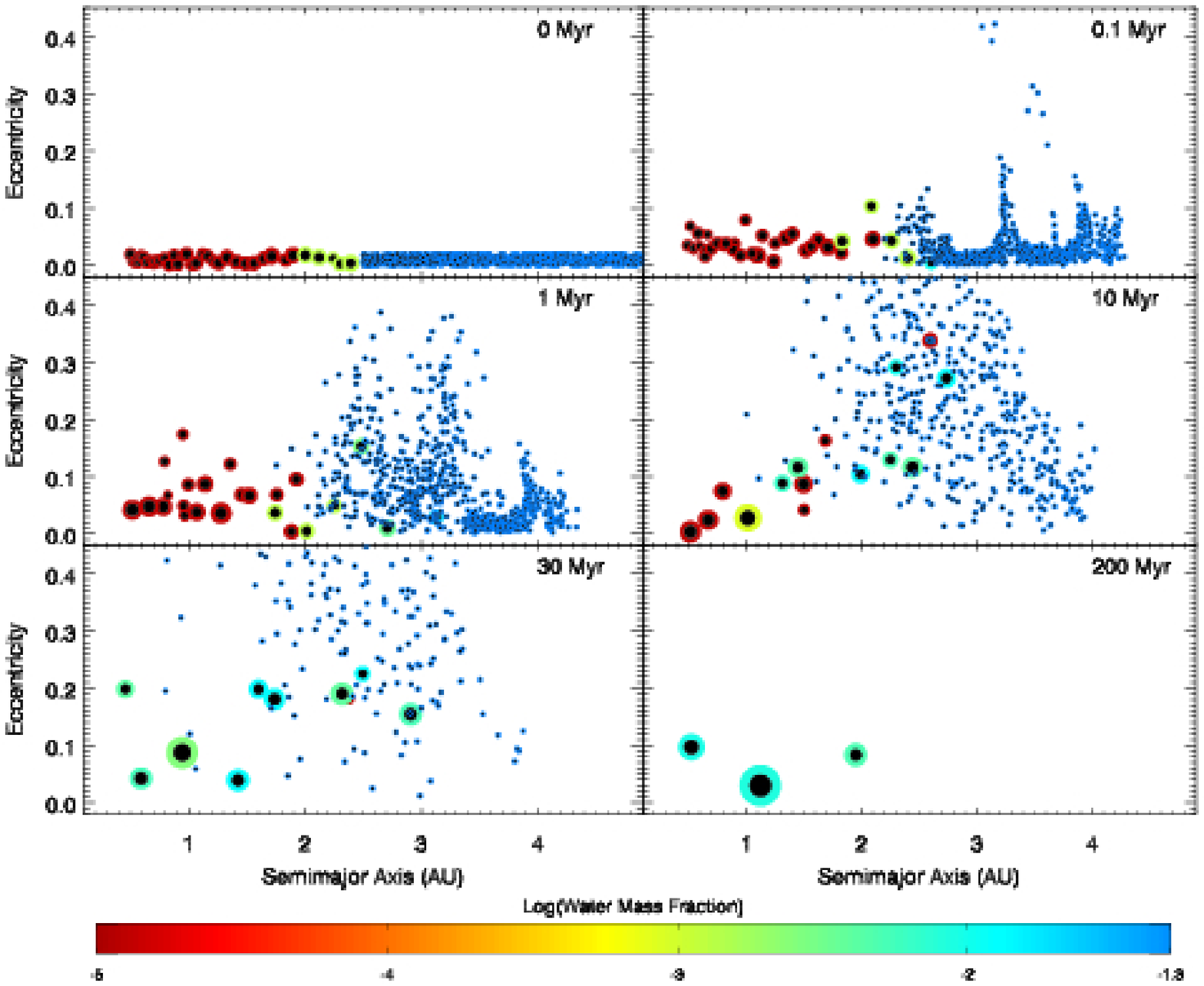,width=12cm}}
\caption{Six snapshots in the evolution of simulation 1b, with 1038
  initial particles -- 38 planetary embryos and 1000 non
  self-interacting planetesimals.}
\label{fig:aetfe1b}
\end{figure}

\begin{figure}
\centerline{\psfig{file=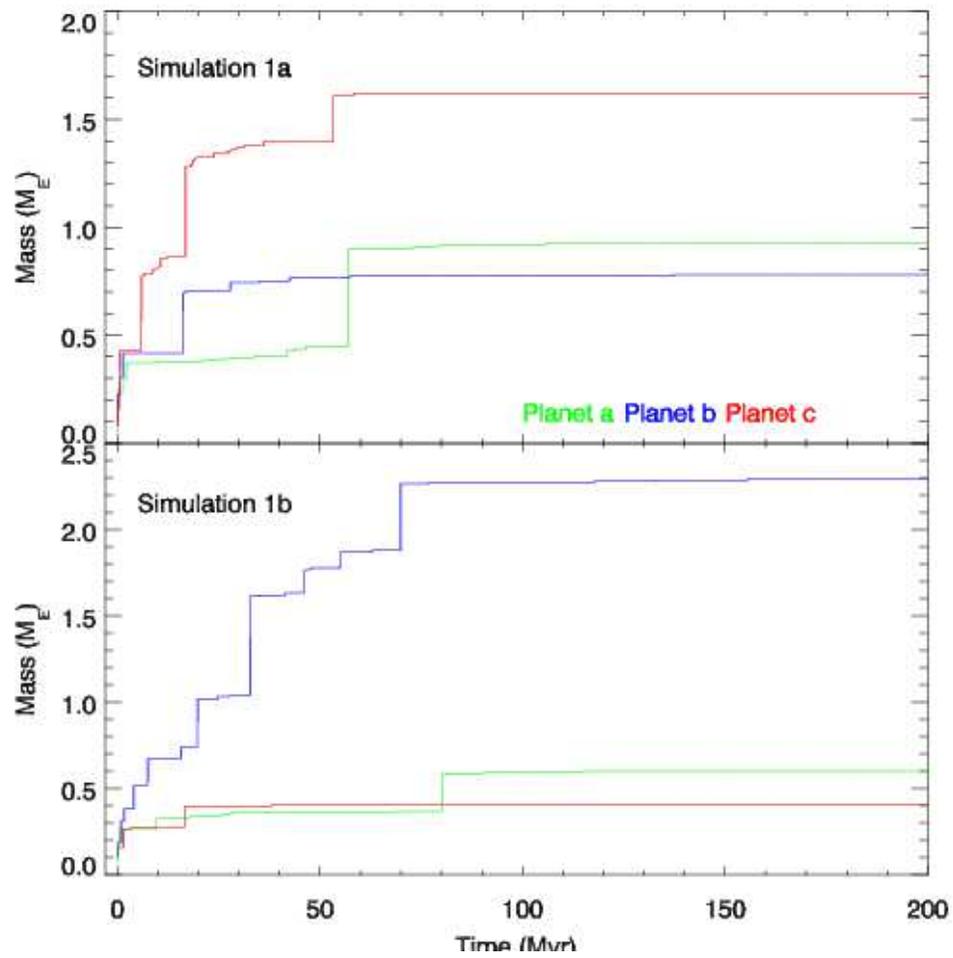,width=13cm}}
\caption{The masses of all surviving bodies from simulations 1a (top
  panel) and 1b (bottom panel) as a function of time.  The innermost
  planet in each case (planet $a$) is shown in green, the middle planet
  (planet $b$) in blue, and the outer planet (planet $c$) in red.}
\label{fig:massvs1}
\end{figure}

\begin{figure}
\centerline{\psfig{file=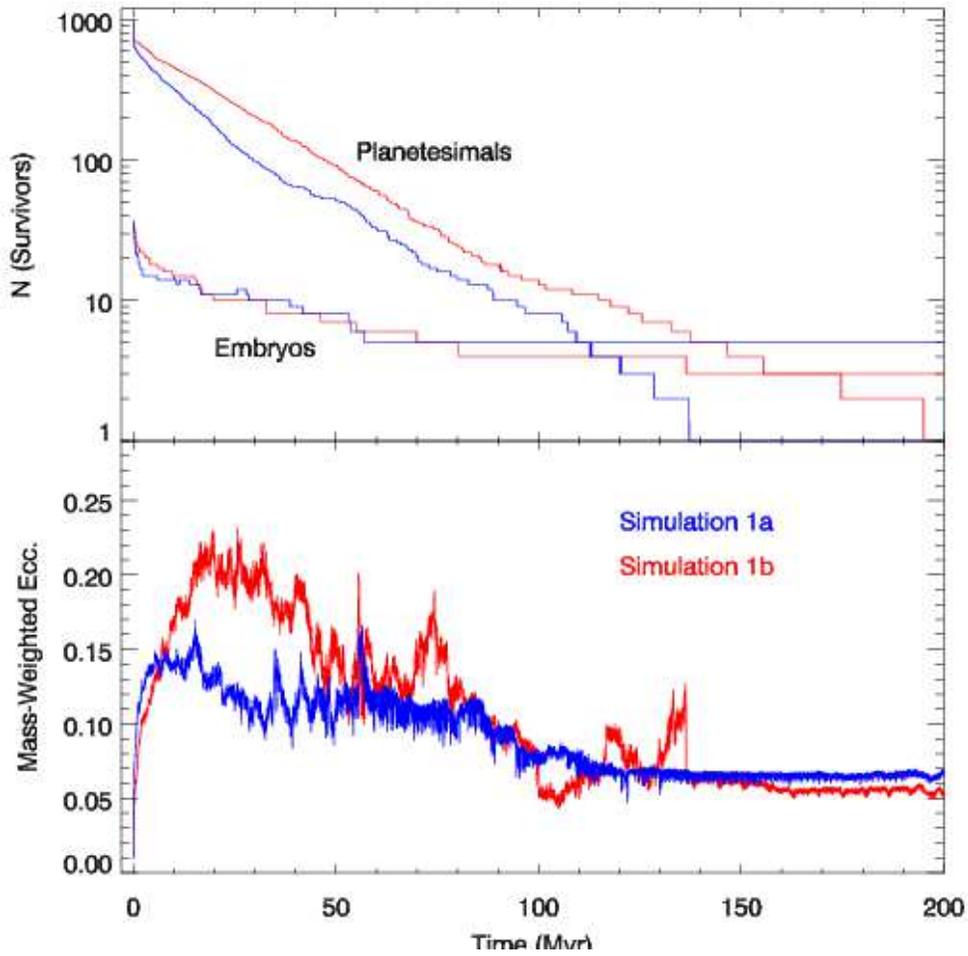,width=13cm}}
\caption{A comparison of the evolution of physical properties of
  simulations 1a (shown in blue) and 1b (red).  \underline{Top}: The
  number of surviving bodies, both planetesimals and planetary
  embryos. \underline{Bottom}: The mass-weighted eccentricity of all
  surviving bodies.}
\label{fig:numvs1}
\end{figure}

\begin{figure}
\centerline{\psfig{file=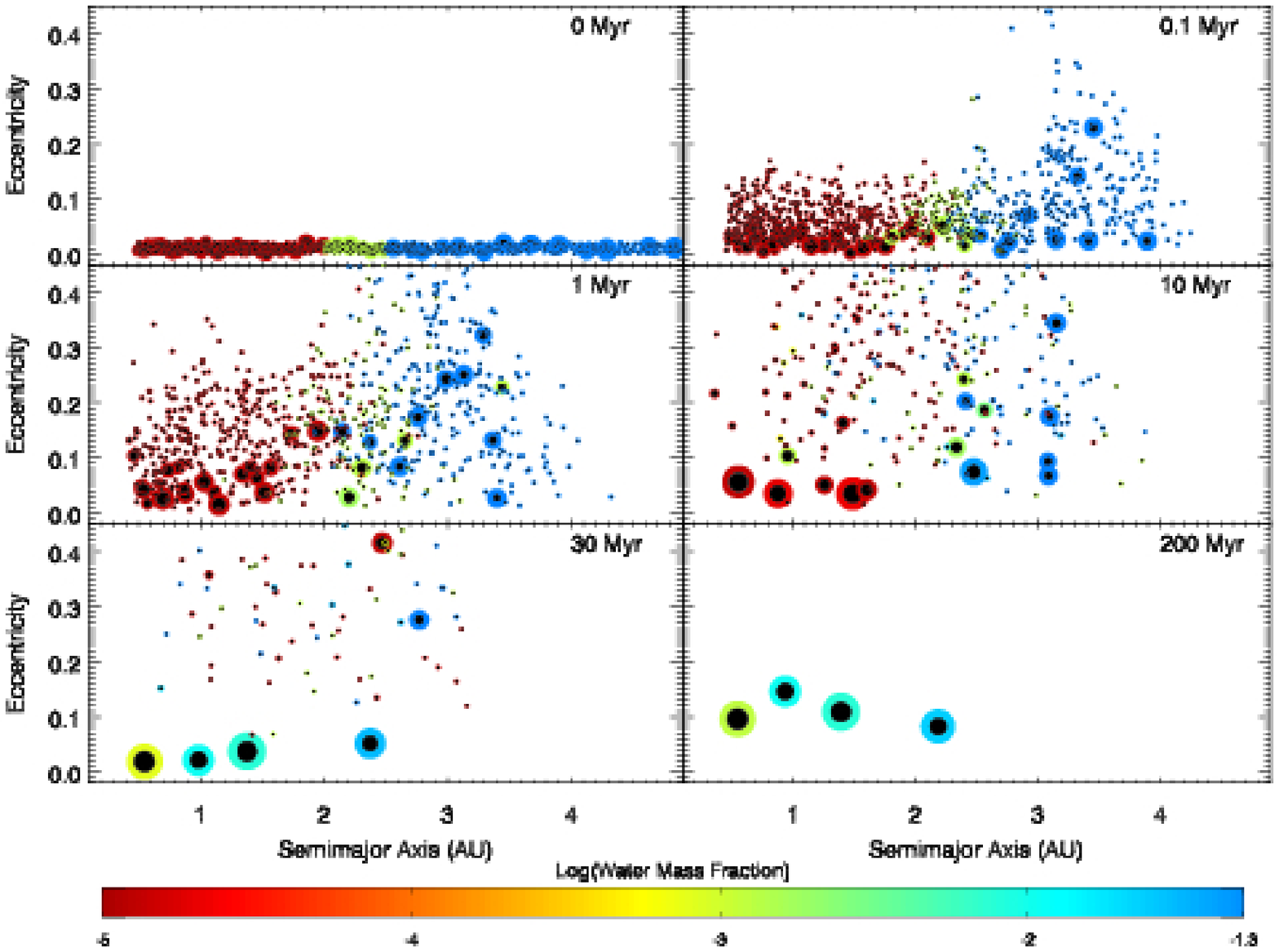,width=12cm}}
\caption{Six snapshots in the evolution of simulation 2a, with 1054
  initial particles, all self-interacting.}
\label{fig:aetfe2a}
\end{figure}

\begin{figure}
\centerline{\psfig{file=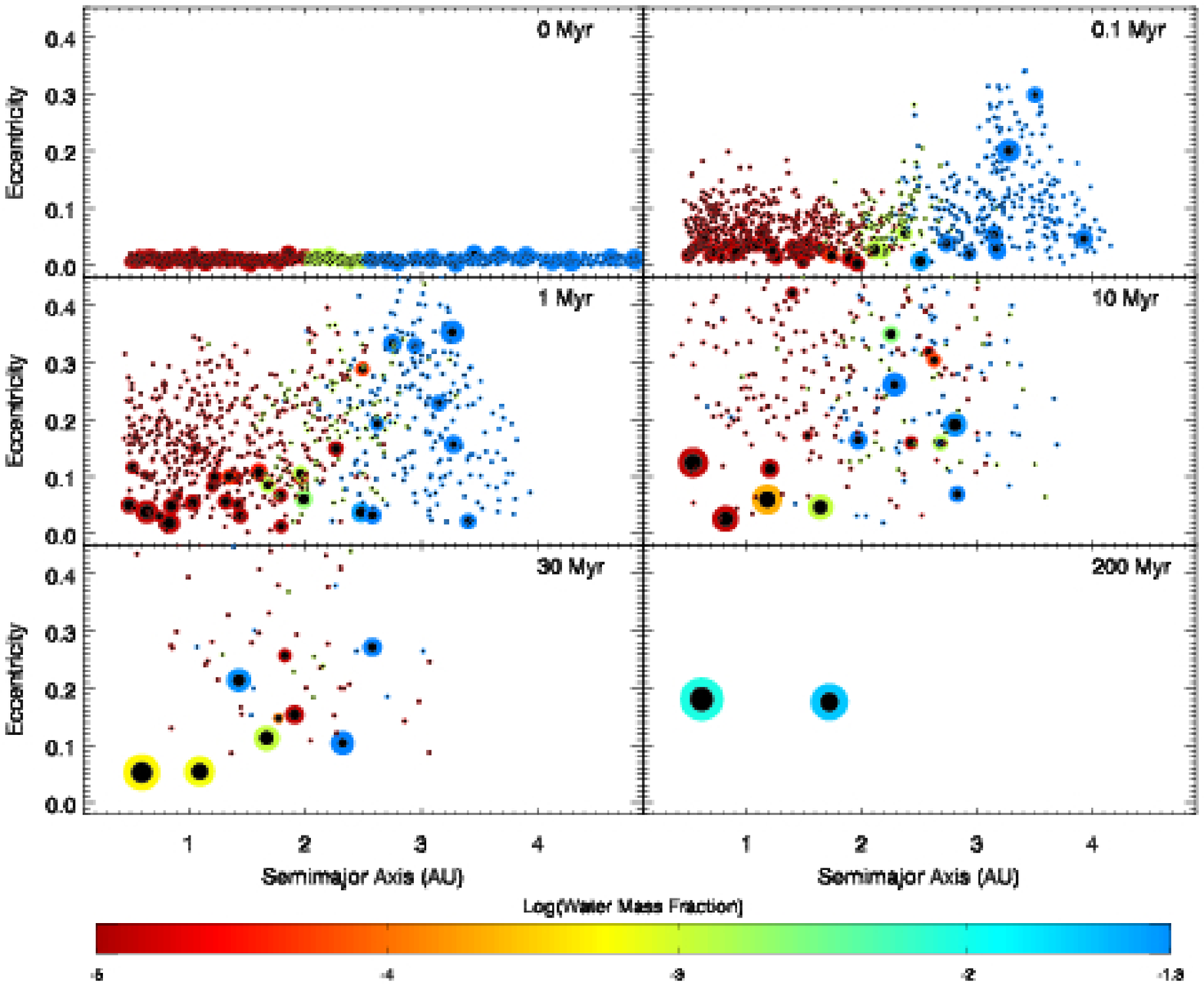,width=12cm}}
\caption{Six snapshots in the evolution of simulation 2b, with 1054
  initial particles -- 54 planetary embryos and 1000 non
  self-interacting planetesimals.}
\label{fig:aetfe2b}
\end{figure}

\begin{figure}
\centerline{\psfig{file=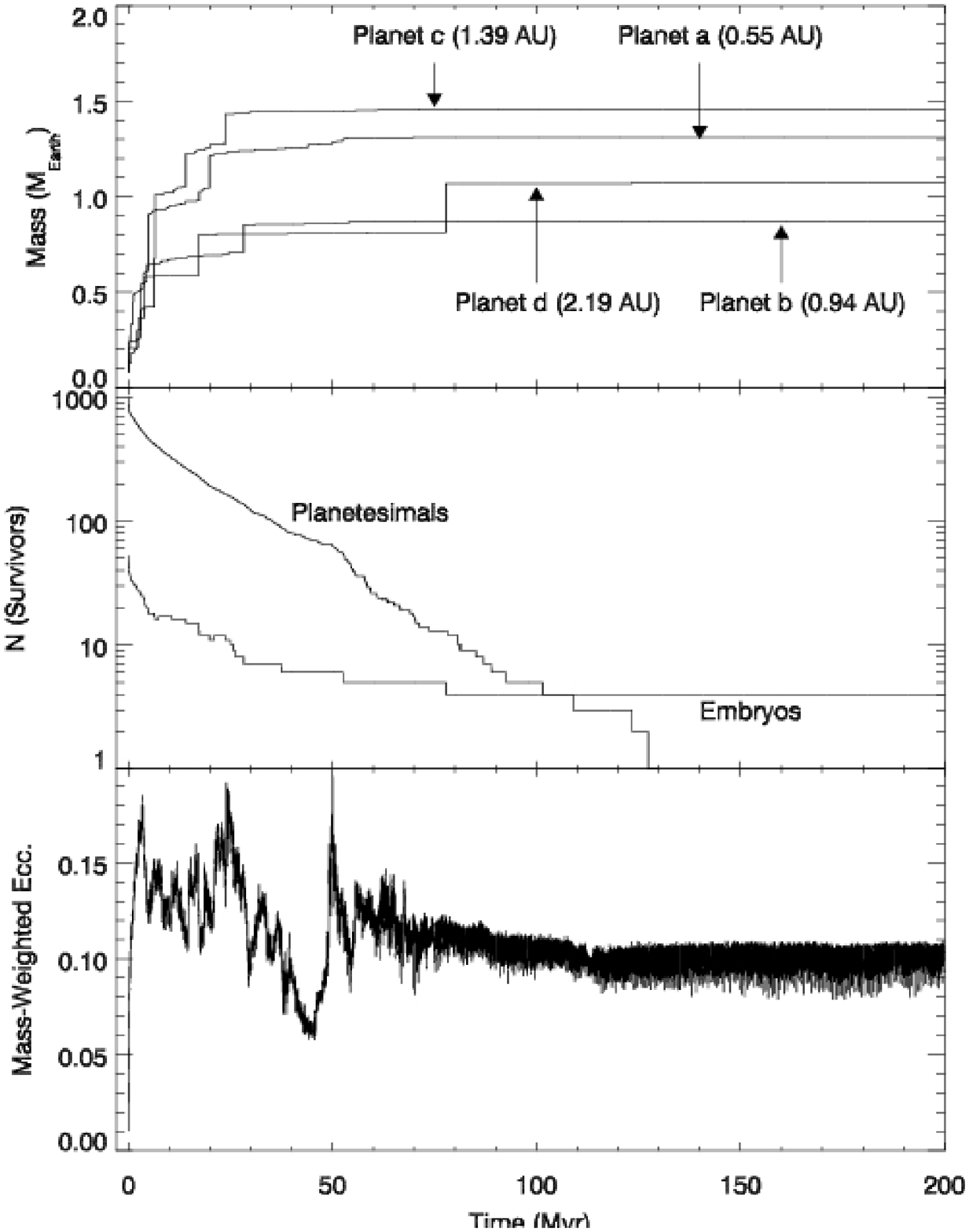,width=13cm}}
\caption{The evolution of simulation 2a.  \underline{Top}: Mass
  vs. time for the four surviving planets. \underline{Middle}: Number
  of surviving planetary embryos through time. \underline{Bottom}:
  Mass-weighted eccentricity as a function of time for all surviving
  bodies in the simulation (excluding Jupiter). }
\label{fig:evol2a}
\end{figure}

\begin{figure}
\centerline{\psfig{file=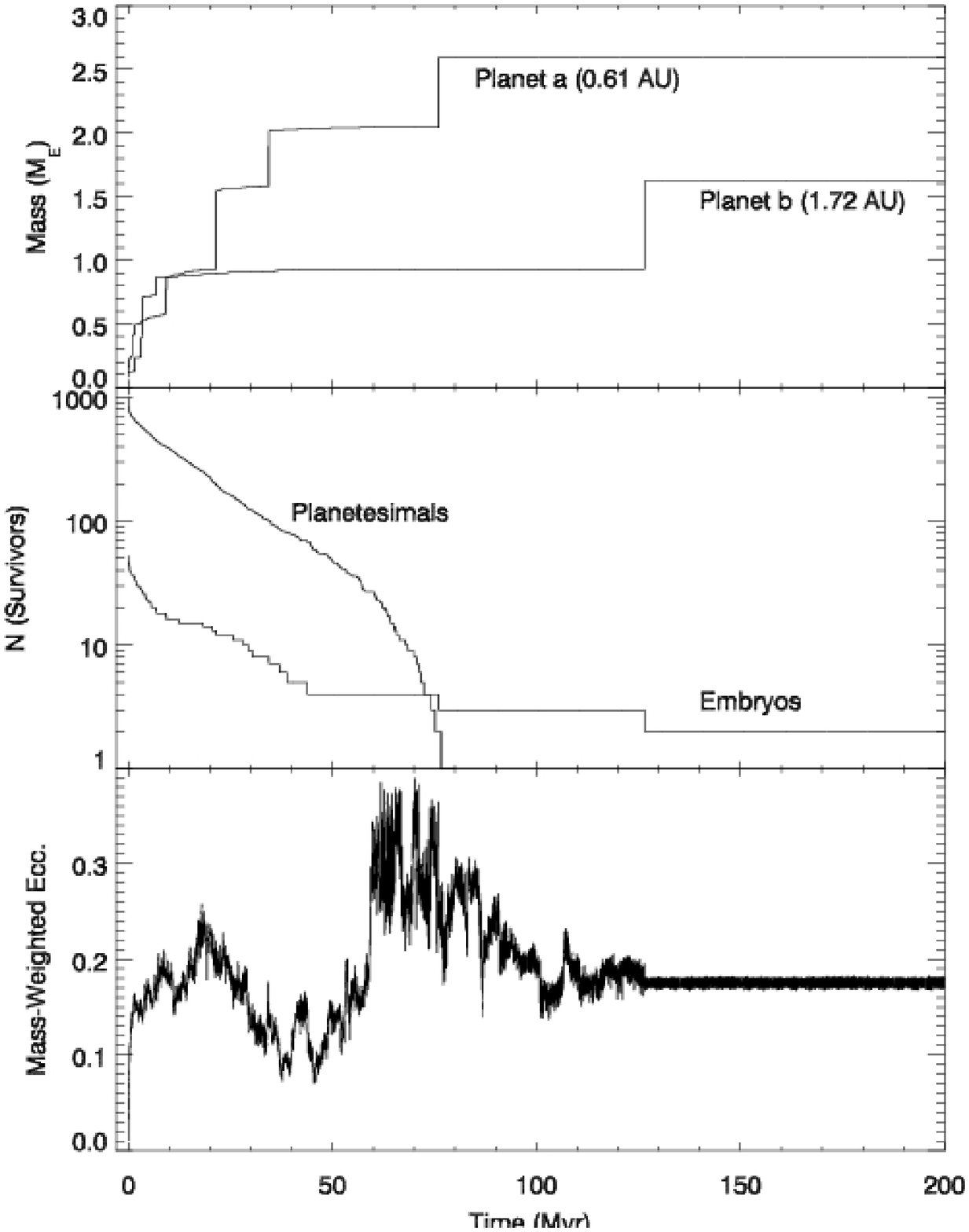,width=13cm}}
\caption{The evolution of simulation 2b.  \underline{Top}: Mass
  vs. time for the two surviving planets. \underline{Middle}: Number
  of surviving planetary embryos through time. \underline{Bottom}:
  Mass-weighted eccentricity as a function of time for all surviving
  bodies in the simulation (excluding Jupiter). }
\label{fig:evol2b}
\end{figure}

We have uncovered several significant, previously unrecognized aspects
of the accretion process:

\begin{enumerate}

\item The feeding zones of the terrestrial planets widen and move outward in
  time, as shown in Fig.~\ref{fig:mloc-t2000}.  While the mass in a
  given annulus is dominated by small bodies, the damping effects
  effectively isolate that region by suppressing eccentricity growth.
  In time, large bodies form and material is scattered out of the
  region, ending its isolation period.  The isolation time corresponds
  roughly to the oligarchic growth phase, and proceeds rapidly in the
  inner disk and moves outward in time.  Assuming that embryos form
  slowly (as in Kokubo \& Ida 2000), planets do not accrete water-rich
  material until at least 10 Myr after the start of a simulation, at
  which time most planets are a sizable fraction of their final mass.

\item 
The growth of planetary embryos in the asteroid belt may have been
stunted by the presence of a giant planet.  If embryos formed
relatively slowly, following the results of Kokubo \& Ida (2000), then
perhaps ten or so smallish, $\sim 0.05 \mearth$ embryos could form in
the inner belt, out to about 3-3.5 AU.  Beyond this, Jupiter's
gravitational effects, as well as intrusion from embryos formed closer
to the Sun, prevented their accretion.  This result is independent of
Jupiter's time of formation, because the 10 Myr timescale for embryos
to form out to 2.5 AU (Kokubo \& Ida 2000) corresponds to the upper
limit on the formation time of Jupiter (Brice\~no et al 2001).
Embryos can form interior to 2.5 AU, be scattered into the asteroid
belt and reside there for long periods of time, but are unlikely to
form {\it in situ}.  However, this result is model-dependent:
Goldreich \etal (2004) calculate much shorter embryo formation
timescales.  In their model, embryos would certainly have formed
before the giant planets, assuming these to form via core-accretion
(Pollack \etal 1996) on a timescale of millions of years.  The speed
of embryo formation has important implications for the robustness of
water delivery: if embryos form quickly, then the number of bodies
involved in water delivery is relatively small, and the process is
relatively stochastic (although less than in previous models --
Morbidelli \etal 2004, RQL04).

\item 
Dynamical friction is a significant phenomenon for the growth of
terrestrial planets.  Indeed, even mall bodies feel dynamical friction
from a swarm of other small bodies.  The high computational cost of
fully-interacting simulations is therefore necessary to ensure a
realistic outcome.  The dynamics and outcome of simulations in which
all particles interacted with each other were different than
simulations in which our ``planetesimals'' were not self-interacting.
Most severe was the case of simulations 2a and 2b.  In simulation 2a,
more planets formed than in simulation 2b (4 planets vs. 2) and the
eccentricities of the final planets were far lower ($e
\, \approx \, 0.1$ vs. $e \, \approx \, 0.2$).  The lifetime of
planetesimals was much longer in simulation 2a, because of dynamical
friction due other small bodies.  The consequences of are important:
the fraction of water delivered to terrestrial planets in the form of
small bodies was twice as high in simulation 2a, and the timing of the
delivery of water-rich embryos and planetesimals also varied among the
two simulations.

\item The asteroid belt is cleared of $>$99\% of its mass as a
natural result of terrestrial accretion.  Embryos and planetesimals
scatter each other onto unstable orbits such as mean motion resonances
with Jupiter.  In time these bodies are removed via ejections, or by
colliding with bodies closer to the Sun.  After $\sim$ 1 billion
years, none of our simulations had more than 0.4\% of their initial
mass remaining in the asteroid belt.  This result was also shown by
Morbidelli \etal (2000) in the context of the current orbits of
Jupiter and Saturn.  In that case, many resonances exist in the
asteroid belt, including the important $\nu_6$ secular resonance with
Saturn at 2.1 AU.  Here we have shown that the asteroid belt is
cleared even in a much calmer dynamical environment, with only one
giant planet on a circular orbit.

\end{enumerate}


\newpage
 
\end{document}